\documentclass[aps,pre,showpacs,twocolumn]{revtex4}
\usepackage{bm}
\usepackage{graphicx}
\bibstyle{approve.bib}
\usepackage{amssymb}
\usepackage{amsmath}
\usepackage{esint}
\usepackage{epstopdf}
\usepackage{color}
\usepackage{gensymb}
\usepackage{mathtools}
\usepackage{suffix}

\newcommand{\be}{\begin{equation}}
\newcommand{\ee}{\end{equation}}
\newcommand{\bea}{\begin{eqnarray}}
\newcommand{\eea}{\end{eqnarray}}
\newcommand{\lan}{\left\langle}
\newcommand{\ran}{\right\rangle}
\newcommand{\br}{\mathbf{r}}
\newcommand{\brc}{\mathbf{r}_{\rm c}}

\newcommand{\bE}{\mathbf{E}}

\newcommand{\bq}{\mathbf{q}}

\newcommand{\bk}{\mathbf{k}}

\newcommand{\e}{\varepsilon}
\newcommand{\rans}{\right\rangle_{\rm s}}

\newcommand{\tG}{\tilde{G}}

\newcommand{\ce}{_{\rm c}}
\newcommand{\s}{_{\rm s}}
\newcommand{\hn}{\hat{n}}
\newcommand{\B}{_{\rm b}}

\begin{document}

\title{Schwinger-Dyson equations for composite electrolytes governed by mixed electrostatic couplings strengths}

\author{Sahin Buyukdagli}
\address{Department of Physics, Bilkent University, Ankara 06800, Turkey}

\begin{abstract}
The electrostatic Schwinger-Dyson equations are derived and solved for an electrolyte mixture composed of mono- and multivalent ions confined to a negatively charged nanoslit. The closure of these equations is based on an asymmetric treatment of the ionic species with respect to their electrostatic coupling strength; the weakly coupled monovalent ions are treated within a gaussian approximation while the multivalent counterions of high coupling strength are incorporated with a strong-coupling approach. The resulting self-consistent formalism includes explicitly the interactions of the multivalent counterions with the monovalent salt. In highly charged membranes characterized by a pronounced multivalent counterion adsorption, these interactions take over the salt-membrane charge coupling. As a result, the increment of the negative membrane charge brings further salt anions into the pore and excludes salt cations from the pore into the reservoir. The corresponding like-charge attraction and opposite-charge repulsion effect is amplified by the pore confinement but suppressed by salt addition into the reservoir. The effect is particularly pronounced in high dielectric membranes where the attractive polarization forces lead to a dense multivalent cation layer at the membrane walls. These cation layers act as an effective positive surface charge, resulting in a total monovalent cation exclusion and a strong anion excess even in the case of neutral membrane walls.
\end{abstract}


\date{\today}
\maketitle   

\section{Introduction}

Electrostatic interactions occurring in aqueous salt solutions are the universal regulators of the biological mechanisms driving life on Earth. From the dense packing of charged biomolecules in confined cell environment to gene expression and viral infection, these interactions govern various in and out of equilibrium processes present in living organisms~\cite{biomatter}. Our correct understanding and control of the biological systems thus necessitates the accurate modeling of the electrostatic coupling between their building blocks. This objective continues to motivate intensive theoretical research work.

The quantitatively accurate description of electrostatic interactions in electrolytes has been initiated by the introduction of the Poisson-Boltzmann (PB) formalism by Gouy~\cite{Gouy} and Chapman~\cite{Chapman} a century ago. The underlying principle behind this formalism consists of the asymmetric treatment of the fixed macromolecular charge density  $\rho\s(\br)$ and the mobile salt charge density $\rho\ce(\br)$. More precisely, given a dielectric permittivity profile $\e(\br)$, the PB theory upgrades the electrostatic Poisson equation 
\be
\label{po}
\nabla\cdot\epsilon(\br)\nabla V(\br)=-\rho\s(\br)-qe\rho\ce(\br)
\ee
by imposing to the mobile ions coupled to the background potential $V(\br)$ a Boltzmann distribution of the form
\be\label{dmf}
\rho\ce(\br)\propto e^{-qeV(\br)/(k_{\rm B}T)},
\ee
with the ionic valency $q$ and electron charge $e$, and the thermal energy $k_{\rm B}T$. The solution of the non-linear PB Eq.~(\ref{po}) provides self-consistently the electric field $\bE(\br)=-\nabla V(\br)$, the salt density $\rho\ce(\br)$, and various related thermodynamic functions such as the surface tension and intermolecular interaction energy profile required for the determination of the stability conditions in biophysical systems.

Despite its major contribution to our understanding of living matter, the PB Eq.~(\ref{po}) presents a critical limitation; the unique type of electrostatic interactions taken into account by this formalism consists of the coupling between the membrane charge and the salt ions, while the theory neglects the energetic contribution from the explicit ion-ion interactions to the salt charge density~(\ref{dmf}). As confirmed by alternative derivations of the PB Eq.~(\ref{po}) from more accurate theories~\cite{PodWKB,netzcoun}, this approximation corresponds to a mean-field (MF) description of inhomogeneous charged liquids valid for weakly charged macromolecules, salt solutions of single valency and high concentration, and dielectrically uniform systems. In the opposite regime of dilute or multivalent salt solutions, and strongly charged macromolecules such as DNA, the emergence of charge correlations leads to unconventional  behavior inaccessible by the PB formalism, such as the aggregation of like-charged macromolecules~\cite{exp2,exp3,Levin1,Levin2}, the electrophoretic motion of anionic polyelectrolytes along the applied electric field~\cite{Qiu2015}, and anionic streaming current through negatively charged pores~\cite{Heyden2006}.

The investigation of charge correlations in inhomogeneous electrolytes was initiated by Wagner's identification of repulsive image-charge forces as the underlying mechanism behind the experimentally measured surface tension excess of salt solutions~\cite{wagner}. This pioneering work was subsequently extended by Onsager and Samaras who derived a limiting law for the electrolyte surface tension in the regime of low salt concentration~\cite{onsager}. Being limited to the regime of weak electrostatic potential fluctuations strictly valid for neutral interfaces, the aforementioned theories are not adequate for describing charge correlation effects in macromolecular interactions where strong surface charges are commonly involved.

The first theoretical description of one-loop (1l) level correlation effects on the interaction of charged membranes has been developed by Podgornik and Zeks within a field-theoretic formulation of inhomogeneous electrolytes~\cite{PodWKB} and by Attard et al. via the solution of a modified Poisson-Boltzmann (PB) equation~\cite{attard}. The former work predicted the correlation-driven attraction between similarly charged membranes previously observed in numerical simulations~\cite{Sim1}. Then, the role played by 1l charge correlations on the interfacial ion densities has been investigated for counterion liquids by Netz and Orland~\cite{netzcoun} and for a salt solution distributed symmetrically around a thin charged interface by Lau~\cite{1loop}. In Ref.~\cite{Buyuk2012}, we solved the 1l-level equations of state and characterized charge fluctuation effects in the experimentally relevant case of a salt solution in contact with a charged dielectric membrane impenetrable to ions. We also extended this formalism to the cylindrical nanopore geometry in Ref.~\cite{BuyukCyl}. In these works, we developed as well a truncated solution of the weak-coupling (WC) variational equations derived by Netz and Orland~\cite{Netz2003} in the presence of macromolecular charges of arbitrary strength. An upgraded version of the Onsager-Samaras theory for neutral interfaces has been also developed within a variational optimization scheme by Hatlo and Lue~\cite{hatlo}.

In a counterion liquid of ionic valency $q$ in contact with a hard wall of surface charge density $\sigma\s$, the weight of the correlations responsible for the departure from the MF behavior is quantified by the electrostatic coupling parameter $\Xi=2\pi q^3\ell_{\rm B}^2\sigma\s$ where $\ell_{\rm B}\approx7$ {\AA} stands for the Bjerrum length~\cite{netzcoun}. Being based on the expansion of the liquid grand potential in terms of the coupling parameter $\Xi$ scaling cubically with the ion valency, the validity of the 1l approach is limited to monovalent liquids ($q=1$) typically located in the weak-coupling (WC) electrostatic regime $\Xi\lesssim 1$. This indicates the inadequacy of the 1l approach for the treatment of multivalent solutions where the coupling parameter can  reach the domain $\Xi\gtrsim100$ defining the electrostatic strong-coupling (SC) regime. 

A critical step towards the quantitative understanding of SC electrostatics has been taken by Moreira and Netz in Ref.~\cite{NetzSC,NetzSC2}. The Authors developed a systematic SC approximation for the evaluation of the grand potential of a counterion liquid interacting with strongly charged macromolecules. Then, Hatlo and Lue developed a self-consistent SC theory~\cite{hatloepl} based on the variational optimization of the splitting of long and short range interactions introduced by Santangelo~\cite{Santangelo}. 

In biological systems,  multivalent counterions usually coexist with background monovalent salt. This complication has been incorporated into the SC electrostatics by Kanduc et al. in Refs.~\cite{Podgornik2010,Podgornik2011}. Namely, the Authors developed a {\it dressed counterion} theory where the background salt was treated at the WC Debye-H\"{u}ckel (DH) level and the additional multivalent counterions were incorporated within the SC approximation. Then, in Ref.~\cite{Buyuk2019}, we upgraded this formalism by treating the monovalent salt interactions at the full 1l-level. Within this {\it SC-dressed 1l theory}, we investigated the polymer adsorption onto like-charged membranes by added multivalent counterions. Finally, the variational SC theory of Ref.~\cite{hatloepl} has been  extended by the Authors via the inclusion of monovalent salt treated at the MF PB level~\cite{hatlosoft}. 

In this article, we carry the perturbative theory of Ref.~\cite{Buyuk2019} to a higher order and develop a self-consistent formalism of mixed electrolytes composed of mono- and multivalent charges. First, in Sec.~\ref{th}, we derive the formally exact Schwinger-Dyson (SD) equations for the charged liquid in contact with arbitrary macromolecular charges. The closure of these equations is chosen according to the coupling strength of the different ionic species in the liquid. Namely, the weakly coupled background monovalent salt is treated within a gaussian approximation that assumes moderate potential fluctuations around the MF PB solution. Then, the multivalent counterions of high coupling strength are considered within a low fugacity expansion equivalent to a SC approximation. 

The resulting strong-coupling SD (SCSD) formalism incorporates the direct effect of the multivalent counterions on the monovalent salt partition in the pore. This explicit many-body effect absent in the previous salt-dressed SC formalisms~\cite{Podgornik2010,Podgornik2011,Buyuk2019} is the key progress of our work.  The electrostatic many-body picture emerging from the theory indicates that the monovalent salt affinity of the pore is set by the competition between the salt-multivalent charge interactions and the salt-membrane charge coupling. In Sec.~\ref{pars}, this hierarchy is fully characterized for charged slit pores in terms of the experimentally accessible model parameters. We find that beyond a characteristic pore charge, the interaction of the strongly adsorbed multivalent counterions with the monovalent salt prevails the salt-membrane charge coupling. As a result, a further rise of the membrane charge enhances the coion density and reduces the counterion density in the pore. We show that the corresponding like-charge attraction and opposite charge repulsion effect is amplified by the pore confinement or a high membrane permittivity but suppressed by added bulk salt. Our results are summarized and the potential applications of the SCSD formalism are discussed in Conclusions.

\section{Theory}
\label{th}

\subsection{Derivation of the liquid partition function}
\label{part}

\begin{figure}
\includegraphics[width=1.\linewidth]{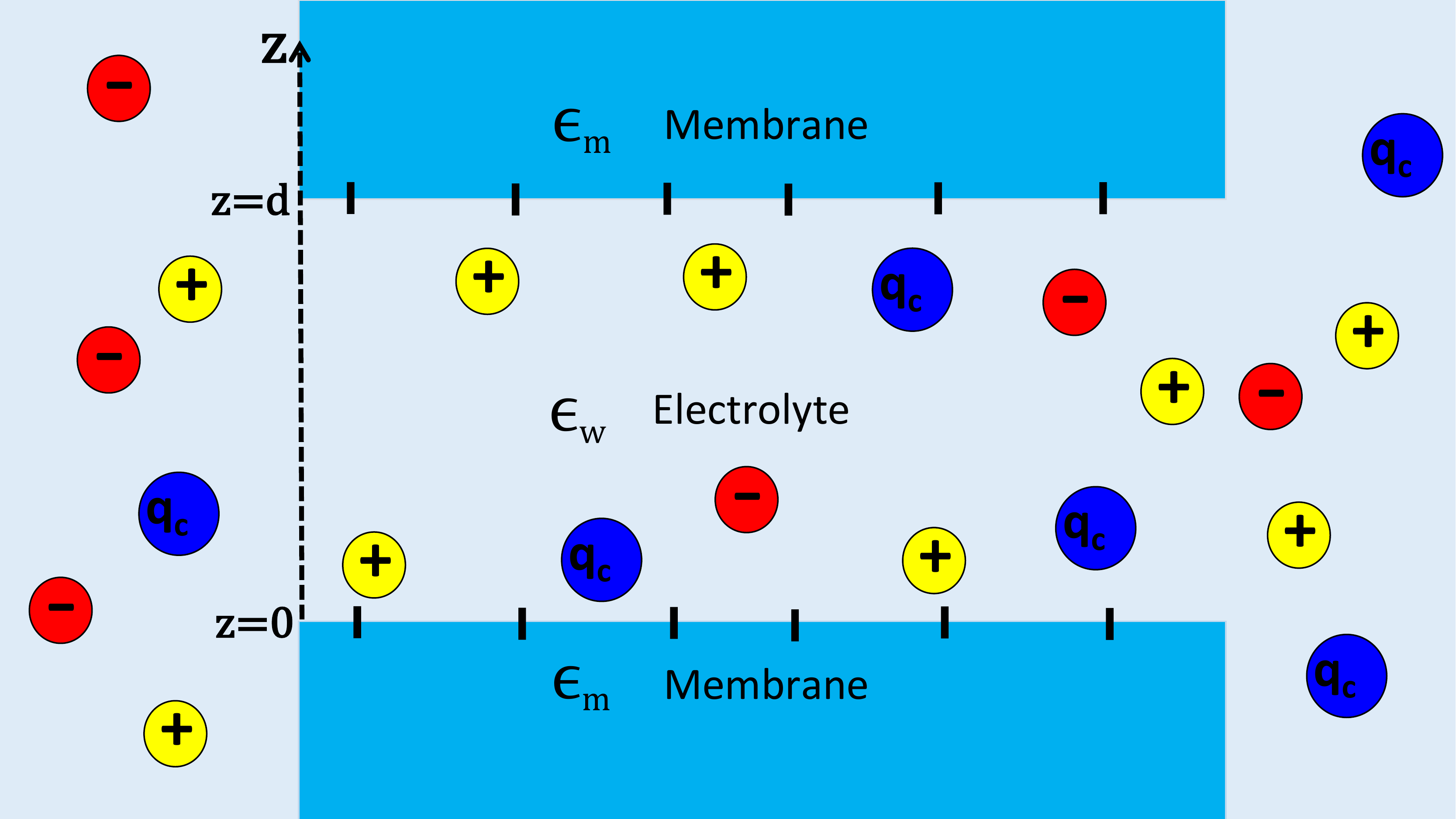}
\caption{(Color online) Schematic depiction of the charged liquid confined to a nanoslit of thickness $d$. At its ends, the slit of negative surface charge density $-\sigma\s$ is in contact with an ionic reservoir containing monovalent cations and anions of respective concentrations $n_{+\rm b}$ and $n_{-\rm b}$, and multivalent counterions of valency $q\ce$ and concentration $n_{\rm cb}$.}
\label{fig1}
\end{figure}

In this part, we derive the partition function of the charged liquid in a functional integral form adequate for analytical treatment~\cite{PodWKB}. The schematic depiction of the charged system is presented in Fig.~\ref{fig1}.  The nanoslit of thickness $d$ and negative surface charge $-\sigma\s$ is located in a solid membrane of dielectric permittivity $\e_{\rm m}$. The slit connected to an external ion reservoir contains an electrolyte of dielectric permittivity $\e_{\rm w}=80$ and temperature $T=300$ K. In addition to implicit solvent, the electrolyte  is composed of $p$ ionic species, with the ionic species $i$ of valency $q_i$ and total number $N_i$. 

The canonical partition function of the confined electrolyte is given by
\bea\label{part}
Z\ce&=&\prod_{i=1}^p\prod_{j=1}^{N_i}\int\mathrm{d}\br_{ij}e^{-\frac{1}{2}\int\mathrm{d}r\hat{\rho}_{\rm c}(\br)v\ce(\br,\br')\hat{\rho}_{\rm c}(\br')-V_i(\br_{ij})}\nonumber\\
&&\hspace{1.85cm}\times e^{-\frac{1}{2}\int\mathrm{d}r\hat{\rho}_{\rm c}(\br)w(\br-\br')\hat{\rho}_{\rm c}(\br')+E\s},
\eea
with the steric potential $V_i(\br_{ij})$ restricting the position of the ion $j$ of the species $i$ to the phase space accessible to the charges, and the charge and particle number densities
\bea\label{dench}
\hat{\rho}\ce(\br)&=&\sum_{i=1}^pq_i\sum_{j=1}^{N_i}\delta(\br-\br_{ij})+\sigma(\br),\\
\label{denn}
\hat{\rho}_{\rm n}(\br)&=&\sum_{i=1}^p\sum_{j=1}^{N_i}\delta(\br-\br_{ij}),
\eea
where the negative membrane surface charge density is
\be
\label{dench2}
\sigma(\br)=-\sigma\s\left[\delta(z)+\delta(d-z)\right].
\ee
In Eq.~(\ref{part}), we used the Coulomb potential defined as the inverse of the operator
\be
v\ce^{-1}(\br,\br')=-\frac{k_{\rm B}T}{e^2}\nabla\cdot\e(\br)\nabla\delta(\br-\br').
\ee
We introduced as well the hard-core (HC) ion-ion interaction potential defined as $w(\br-\br')=\infty$ if $||\br-\br'||\leq2a$ and $w(\br-\br')=0$ for $||\br-\br'||>2a$, where we assumed the same hydrated ion radius $a$ for all species. Finally, we subtracted from the total interaction energy the ionic self-energy $E\s=\sum_{i=1}^pN_i\epsilon_i$ in a pure bulk solvent, with the bulk self-energy of the species $i$ defined as $\epsilon_i=\left[q_i^2v_{\rm cb}(0)+w(0)\right]/2$, and the bulk Coulomb potential $v_{\rm cb}(r)=\ell_{\rm B}/r$ where $\ell_{\rm B}=e^2/(4\pi\ell_{\rm B} k_{\rm B}T)\approx 7$ {\AA} is the Bjerrum length.

Introducing in Eq.~(\ref{part}) an Hubbard-Stratonovich transformation for each type of pairwise interaction potential, the grand-canonical partition function 
\be
\label{zg1}
Z_{\rm G}=\prod_{i=1}^p\sum_{N_i\geq0}\frac{\Lambda_i^{N_i}}{N_i!}Z_{\rm c}
\ee
takes the functional integral form~\cite{NetzHC} 
\be\label{zg2}
Z_{\rm G}=\int\mathcal{D}\phi\mathcal{D}\psi\;e^{-H[\phi,\psi]}, 
\ee
with the Hamiltonian functional 
\bea\label{HamFunc}
H[\phi,\psi]&=&\frac{k_{\rm B}T}{2e^2}\int\mathrm{d}\br\left[\nabla\phi(\br)\right]^2-i\int\mathrm{d}\br\sigma(\br)\phi(\br)\nonumber\\
&&+\frac{1}{2}\int\mathrm{d}\br\mathrm{d}\br'\psi(\br)w^{-1}(\br-\br')\psi(\br')\nonumber\\
&&-\sum_{i=1}^p\Lambda_i \int\mathrm{d}\br\;e^{\epsilon_i-V_i(\br)+iq_i\phi(\br)+i\psi(\br)}
\eea
where $\Lambda_i$ stands for the fugacity of the ionic species $i$. The first three terms on the r.h.s. of Eq.~(\ref{HamFunc}) are respectively the energetic contribution from the solvent, the fixed membrane charges, and the HC interactions to the grand potential. The fourth term corresponds in turn to the contribution from the mobile ions.

The average ion density of the species $i$ follows from the grand potential $\beta\Omega_{\rm G}=-\ln Z_{\rm G}$ as 
\be\label{den1}
n_i(\br)=\frac{\delta\left[\beta\Omega_{\rm G}\right]}{\delta V_i(\br)}=\lan \hn_i(\br)\ran,
\ee
where we introduced the density functional 
\be\label{denf}
\hn_i(\br)=\Lambda_i\;e^{\epsilon_i-V_i(\br)+iq_i\phi(\br)+i\psi(\br)}.
\ee
In Eq.~(\ref{den1}), we used the bracket notation designating the statistical average of a general functional $F[\phi,\psi]$ over the fluctuations of the potentials $\phi(\br)$ and $\psi(\br)$, 
\be
\label{av1}
\lan F[\phi,\psi]\ran=\frac{1}{Z_{\rm G}}\int \mathcal{D}\phi\mathcal{D}\psi\;e^{-H[\phi,\psi]}F[\phi,\psi].
\ee

We now restrict ourselves to the specific case of a solution composed of a monovalent 1:1 salt including cations and anions of valency $q_\pm=\pm1$, fugacity $\Lambda_\pm$, and bulk concentration $n_{\pm\rm b}$, and an additional multivalent cation species of valency $q_{\rm c}$, fugacity $\Lambda_{\rm c}$, and bulk concentration $n_{\rm cb}$. For this electrolyte mixture composed of $p=3$ species, the Hamiltonian functional~(\ref{HamFunc}) reads
\be\label{H02}
H[\phi,\psi]=H_{\rm s}[\phi,\psi]+H_{\rm c}[\phi,\psi], 
\ee
with the Hamiltonian component $H\s[\phi,\psi]$ associated with the solvent, the monovalent salt, and the macromolecular charge, and the multivalent counterion contribution $H\ce[\phi,\psi]$,
\bea
\label{Hs}
H\s[\phi,\psi]&=&\frac{k_{\rm B}T}{2e^2}\int\mathrm{d}\br\;\e(\br)\left[\nabla\phi(\br)\right]^2-i\int\mathrm{d}\br\sigma(\br)\phi(\br)\nonumber\\
&&+\frac{1}{2}\int\mathrm{d}\br\mathrm{d}\br'\psi(\br)w^{-1}(\br-\br')\psi(\br')\nonumber\\
&&-\sum_{i=\pm}\int\mathrm{d}\br\;\hn_i(\br),\\
\label{Hq}
H_{\rm c}[\phi,\psi]&=&-\int\mathrm{d}\br\;\hn\ce(\br).
\eea

\subsection{Derivation of the electrostatic SCSD equations}
\label{scex}

\subsubsection{SC expansion of the SD equations}

We derive here the electrostatic equations of state extending the PB equation beyond the MF regime. The corresponding SCSD equations are intended to account for the WC correlations governing the monovalent salt partition and the SC correlations mediated by the multivalent counterions. In Appendix~\ref{ap1}, we show that the SD equations associated with a functional integral of the form~(\ref{zg2}) are given by the formally exact identities
\bea
\label{SD1}
&&\lan \frac{\delta H[\phi,\psi]}{\delta \phi(\br)}\ran=0,\\
\label{SD2}
&&\lan \frac{\delta H[\phi,\psi]}{\delta \phi(\br)}\phi(\br')\ran=\delta(\br-\br').
\eea
The equalities~(\ref{SD1}) and~(\ref{SD2}) result from the invariance of the partition function with respect to an infinitesimal variation of the electrostatic potential $\phi(\br)$. 

In order to simplify the notation, from now on, we will omit the dependencies of the functionals on the potentials $\phi(\br)$ and $\psi(\br)$. At this point, we introduce the SC approximation equivalent to a cumulant expansion of the partition function~(\ref{zg2}) in terms of the multivalent counterion density~\cite{NetzSC,Podgornik2010}. Namely, by using the decomposition in Eqs.~(\ref{H02})-(\ref{Hq}), we first Taylor-expand Eq.~(\ref{av1}) at the linear order in the counterion density $\hn\ce(\br)$ to get
\be
\label{avex}
\lan F\ran=\lan F\rans+\int\mathrm{d}\brc\left[\lan F\hn\ce(\brc)\rans-\lan F\rans\lan\hn\ce(\brc)\rans\right],
\ee
where we introduced the field average with respect to the salt Hamiltonian in Eq.~(\ref{Hs}),
\be
\label{brs}
\lan F\rans=\frac{1}{Z\s}\int \mathcal{D}\phi\mathcal{D}\psi\;e^{-H\s}F,
\ee
with the salt partition function $Z\s=\int \mathcal{D}\phi\mathcal{D}\psi\;e^{-H\s}$. Then, we use Eq.~(\ref{avex}) to expand the SD Eqs.~(\ref{SD1}) and~(\ref{SD2}) at the same order $O\left(\hn\ce\right)$. This yields the SC-expanded SD (SCSD) equations
\bea
\label{SD3}
&&\lan \frac{\delta H}{\delta \phi(\br)}\rans=\lan \frac{\delta H\s}{\delta \phi(\br)}H\ce\rans-\lan \frac{\delta H\s}{\delta \phi(\br)}\rans\lan H\ce\rans,\\
\label{SD4}
&&\lan \frac{\delta H}{\delta \phi(\br)}\phi(\br')\rans-\lan \frac{\delta H\s}{\delta \phi(\br)}\phi(\br')H\ce\rans\nonumber\\
&&+\lan \frac{\delta H\s}{\delta \phi(\br)}\phi(\br')\rans\lan H\ce\rans=\delta(\br-\br').
\eea
Finally, applying the equality~(\ref{avex}) to Eq.~(\ref{den1}), the SC-expanded mono- and multivalent ion densities become
\bea
\label{de1}
n_\pm(\br)&=&\lan \hn_\pm(\br)\rans+\int\mathrm{d}\brc\left[\lan \hn_\pm(\br)\hn\ce(\brc)\rans\right.\\
&&\left.\hspace{2.8cm}-\lan \hn_\pm(\br)\rans\lan\hn\ce(\brc)\rans\right],\nonumber\\
\label{de2}
n\ce(\br)&=&\lan \hn\ce(\br)\rans.
\eea

In the derivation of Eq.~(\ref{de2}), we dropped a term of order $O\left(\Lambda\ce^2\right)$. We also note that from now on, our derivations will be based on a systematic linearization of the equations in terms of the bulk counterion fugacity $\Lambda\ce$ or equivalently the concentration $n_{\rm cb}$. Substituting now the Hamiltonian functional and its components in Eqs.~(\ref{H02})-(\ref{Hq}) into the SCSD Eqs.~(\ref{SD3}) and~(\ref{SD4}), the latter take the following form,
\begin{widetext}
\bea\label{e1}
&-&\frac{k_{\rm B} T}{e^2}\lan\nabla\e(\br)\nabla\phi(\br)\rans-i\sigma(\br)-i\sum_{i=\pm,{\rm c}}q_i\lan\hn_i(\br)\rans\\
&=&\frac{k_{\rm B} T}{e^2}\int\mathrm{d}\brc\left[\lan\hn_{\rm c}(\brc)\nabla\e(\br)\nabla\phi(\br)\rans-\lan\hn_{\rm c}(\brc)\rans\lan\nabla\e(\br)\nabla\phi(\br)\rans\right]
+i\sum_{i=\pm}q_i\int\mathrm{d}\brc\left[\lan\hn_{\rm c}(\brc)\hn_i(\br)\rans-\lan\hn_{\rm c}(\brc)\rans\lan\hn_i(\br)\rans\right],\nonumber\\
\label{e2}
&-&\frac{k_{\rm B} T}{e^2}\lan\phi(\br')\nabla\e(\br)\nabla\phi(\br)\rans-i\sigma(\br)\lan\phi(\br')\rans-i\sum_{i=\pm,{\rm c}}\lan\hn_i(\br)\phi(\br')\rans-\delta(\br-\br')\\
&=&\frac{k_{\rm B} T}{e^2}\int\mathrm{d}\brc\left[\lan\hn_{\rm c}(\brc)\phi(\br')\nabla\e(\br)\nabla\phi(\br)\rans-\lan\hn_{\rm c}(\brc)\rans\lan\phi(\br')\nabla\e(\br)\nabla\phi(\br)\rans\right]\nonumber\\
&&+i\sum_{i=\pm}q_i\int\mathrm{d}\brc\left[\lan\hn_{\rm c}(\brc)\hn_i(\br)\phi(\br')\rans-\lan\hn_{\rm c}(\brc)\rans\lan\hn_i(\br)\phi(\br')\rans\right]+i\sigma(\br)\int\mathrm{d}\brc\left[\lan\hn_{\rm c}(\brc)\phi(\br')\rans-\lan\hn_{\rm c}(\brc)\rans\lan\phi(\br')\rans\right].\nonumber
\eea
\end{widetext}

The l.h.s. of Eqs.~(\ref{e1}) and~(\ref{e2}) have respectively the form of a PB-like equation for the average electrostatic potential $\lan \phi(\br)\rans$ induced by the fixed charge $\sigma(\br)$, and a screened Laplace equation for the electrostatic correlator $\lan\phi(\br)\phi(\br')\rans$. These equalities are augmented on their r.h.s. by the direct coupling between the mono- and multivalent ion densities. The explicit effect of the multivalent counterions on the monovalent salt partition is precisely embodied in these coupling terms. In Sec.~\ref{cls}, we explain the evaluation of the field averages in Eqs.~(\ref{e1}) and~(\ref{e2}) within a Gaussian closure approximation.

\subsubsection{Gaussian closure of the SCSD Eqs.~(\ref{e1})-(\ref{e2})}
\label{cls}

Due to the non-linearity of the salt Hamiltonian~(\ref{Hs}), the field averages in the SCSD Eqs.~(\ref{e1}) and~(\ref{e2}) cannot be evaluated analytically. We will thus assume small fluctuations of the weakly coupled monovalent salt densities around their MF value and approximate the salt Hamiltonian~(\ref{Hs}) by a Gaussian Hamiltonian,  
\bea\label{gaus}
H_{\rm s}\approx H_0&=&\frac{1}{2}\int_{\br,\br'}\left[\phi-i\phi_{\rm 0}\right]_\br G^{-1}(\br,\br')\left[\phi-i\phi_{\rm 0}\right]_{\br'}\nonumber\\
&&+\frac{1}{2}\int_{\br,\br'}\psi(\br)w^{-1}(\br-\br')\psi(\br').
\eea

In the absence of HC interactions, Eq.~(\ref{gaus}) without the second integral term  would correspond to the electrostatic Hamiltonian of the most general quadratic form~\cite{Netz2003}. In the present case, the HC interactions of our model are approximated in Eq.~(\ref{gaus}) by the bare HC potential $w(\br-\br')$. A quadratic expansion of Eq.~(\ref{Hs}) in terms of the potential $\psi(\br)$ shows that this approximation neglects the renormalization of the naked HC potential $w(\br-\br')$ by ionic excluded-volume effects, and the coupling of these interactions to the electrostatic potential $\phi(\br)$. The corresponding approximation is justified by the results of previous MC simulations and theoretical investigations of Yukawa-type core interactions where it was observed that surface wetting by excluded-volume interactions and the variation of the interfacial charge densities by the finite ion size are negligible in the submolar concentration regime considered in the present work~\cite{Buyuk2011}.

The evaluation of the field averages in Eqs.~(\ref{de1})-(\ref{e2}) with the gaussian Hamiltonian~(\ref{gaus}) will be based on the use of the generating functional
\bea\label{gen}
I[J_1,J_2]&=&\lan e^{\int\mathrm{d}\br\left[J_1(\br)\phi(\br)+J_2(\br)\psi(\br)\right]}\ran\s\\
&=&e^{\frac{1}{2}\int_{\br,\br'}J_1(\br)G(\br,\br')J_1(\br')+i\int_\br J_1(\br)\phi_0(\br)}\nonumber\\
&&\times e^{\frac{1}{2}\int_{\br,\br'}J_2(\br)w(\br-\br')J_2(\br')}\nonumber
\eea
and its derivatives with respect to the generating functions $J_{1,2}(\br)$. First, from Eq.~(\ref{gen}), the average value of the electrostatic potential and its variance associated with the monovalent salt follow as 
\bea\label{mean}
\lan\phi(\br)\rans&=&i\phi_0(\br),\\
\label{var}
\lan\phi(\br)\phi(\br')\rans&=&G(\br,\br')-\phi_0(\br)\phi_0(\br').
\eea
Then, using Eqs.~(\ref{avex}) and~(\ref{gen}), the ion densities~(\ref{de1}) and~(\ref{de2}) take the explicit form
\bea\label{deni}
n_\pm(\br)&=&\rho_\pm(\br)\left\{1+\int\mathrm{d}\brc n\ce(\brc)f_\pm(\br,\brc)\right\},\\
\label{denc}
n\ce(\brc)&=&\rho\ce(\brc),
\eea 
where we introduced the naked ion density function 
\be
\label{denibare}
\rho_i(\br)=\Lambda_i\;e^{\frac{q_i^2}{2}v_{\rm cb}(0)-V_i(\br)-q_i\phi_0(\br)-\frac{q_i^2}{2}G(\br,\br)}
\ee
free of the explicit multivalent counterion contribution, and the Mayer function
\be
\label{Mayer}
f_i(\br,\brc)=e^{- q_iq\ce G(\br,\brc)-w(\br-\brc)}-1.
\ee
 
We evaluate now the field averages in the SCSD Eqs.~(\ref{e1}) and~(\ref{e2}) within the gaussian closure approximation of Eq.~(\ref{gaus}). Using extensively the identity~(\ref{gen}), after long and tedious algebraic manipulations, the SCSD Eqs.~(\ref{e1})-(\ref{e2}) finally take the explicit form
\begin{widetext}
\bea
\label{SD5}
&&\frac{k_{\rm B}T}{e^2}\nabla_\br\cdot\e(\br)\nabla_\br\left\{\phi_0(\br)+q\ce\int\mathrm{d}\brc n\ce(\brc)G(\brc,\br)\right\}+\sum_{i=\pm,c}q_in_i(\br)+\sigma(\br)=0,\\
\label{SD6}
&&\left\{\frac{k_{\rm B}T}{e^2}\nabla_\br\cdot\e(\br)\nabla_\br-\sum_{i=\pm,c}q_i^2n_i(\br)\right\}G(\br,\br')-q\ce\int\mathrm{d}\brc n\ce(\brc)F(\br,\brc)G(\brc,\br')=-\delta(\br-\br').
\eea
In Eq.~(\ref{SD6}), we introduced the charge structure function
\be\label{nlsc}
F(\br,\brc)=\frac{k_{\rm B}T}{e^2}\nabla_\br\cdot\e(\br)\nabla_\br\left[\phi_0(\br)+q\ce G(\br,\brc)\right]+\sigma(\br)+\sum_{i=\pm}q_i\rho_i(\br)\left[1+f_i(\br,\brc)\right].
\ee
\end{widetext}

\subsection{Simplifying the SCSD Eqs.~(\ref{SD5})-(\ref{SD6})}
\label{solsc}

In this work, electrostatic correlations will be investigated within the submolar ion concentration regime where the ion size is negligible~\cite{Buyuk2011}. Therefore, we will simplify the SCSD Eqs.~(\ref{SD5})-(\ref{SD6}) by removing the HC interactions. In order to introduce this simplification, we first replace the Laplacian terms in Eq.~(\ref{nlsc}) by the first terms of Eqs.~(\ref{SD5})-(\ref{SD6}), and expand the former at the order $O\left(\Lambda\ce^0\right)$ to obtain~\cite{foot1}
\bea\label{nlsc2}
F(\br,\brc)&=&\sum_{i=\pm}q_i\rho_i(\br)\left[f_i(\br,\brc)+q_iq\ce G(\br,\brc)\right.\nonumber\\
&&\left.\hspace{1.7cm}-q\ce\delta(\br-\brc)\right].
\eea
Next, we set the HC interactions to zero, $w(\br-\br')=0$, and approximate the Mayer function~(\ref{Mayer}) by its Taylor expansion, i.e. $f_i(\br,\brc)\approx- q_iq\ce G(\br,\brc)$.  As this approximation cancels the first line of Eq.~(\ref{nlsc2}), the SCSD Eqs.~(\ref{SD5}) and~(\ref{SD6}) finally reduce to
\bea
\label{SD15II}
&&\frac{k_{\rm B}T}{e^2}\nabla_\br\cdot\e(\br)\nabla_\br\phi_0(\br)+\sum_{i=\pm}q_i\rho_i(\br)+\sigma(\br)=0,\\
\label{SD16II}
&&\left\{\frac{k_{\rm B}T}{e^2}\nabla_\br\cdot\e(\br)\nabla_\br-\sum_{i=\pm}q_i^2n_i(\br)\right\}G(\br,\br')=-\delta(\br-\br'),\nonumber\\
\eea
and the monovalent ion densities in Eq.~(\ref{deni}) become
\be\label{den2}
n_\pm(\br)=\rho_\pm(\br)\left\{1\mp q\ce\int\mathrm{d}\brc n\ce(\brc)G(\br,\brc)\right\}.
\ee

We notice that the simplified SCSD Eqs.~(\ref{SD15II})-(\ref{SD16II}) are similar in form to the WC variational equations of Ref.~\cite{Buyuk2012}. The crucial difference between these two formalisms stems from the presence of the integral term in the ion densities~(\ref{den2}).  In Sec.~\ref{quin}, we discuss qualitatively the physical consequence of this convolution integral incorporating into the SCSD Eqs.~(\ref{SD15II})-(\ref{SD16II}) the SC electrostatics of the multivalent counterions.

\subsection{Effect of the SC electrostatics on monovalent salt partition}
\label{quin}

As noted above, the monovalent ion densities in Eq.~(\ref{den2}) differ from their WC counterpart by the presence of the convolution integral~\cite{Netz2003,Buyuk2012} .  With the aim to understand the origin of this integral term, we use Eq.~(\ref{avex}) to obtain the total average potential as 
\be
\label{mn}
\lan\phi(\br)\ran=i\phi_0(\br)+iq\ce\int\mathrm{d}\brc n\ce(\brc)G(\brc,\br).
\ee
According to Eq.~(\ref{mn}), the net average potential in the liquid is given by the superposition of the salt-dressed potential $\phi_0(\br)$ of Eq.~(\ref{SD15II}) originating from the membrane charges $\sigma(\br)$, and the convolution integral corresponding to the potential induced by all multivalent counterions in the liquid at the specific point $\br$. This implies that the monovalent ion partition characterized by Eq.~(\ref{den2}) is governed by the competition between the salt-membrane interactions embodied by the term $\rho_{\pm}(\br)$ depending on the potential $\phi_0(\br)$, and the salt-multivalent counterion interactions incorporated by the additional convolution integral. This direct influence of the strongly coupled multivalent counterions on the monovalent salt partition is a newly emerging effect that has not been covered by the previous perturbative SC theories~\cite{Podgornik2010,Podgornik2011,Buyuk2019}. Thus, the self-consistent incorporation of the salt-counterion interactions into the SC electrostatics is the key progress of our work. In the limit $\rho_\pm(\br)=0$ where this coupling vanishes, Eqs.~(\ref{SD15II})-(\ref{SD16II}) indeed reduce to the Poisson and Laplace equations in a pure solvent, i.e.
\bea
\label{SD19}
&&\frac{k_{\rm B}T}{e^2}\nabla_\br\cdot\e(\br)\nabla_\br\phi_0(\br)+\sigma(\br)=0,\\
\label{SD20}
&&-\frac{k_{\rm B}T}{e^2}\nabla_\br\cdot\e(\br)\nabla_\br G(\br,\br')=\delta(\br-\br').
\eea
Eqs.~(\ref{SD19}) and~(\ref{SD20}) correspond to the asymptotic SC theory of Ref.~\cite{NetzSC}. Thus, in the absence of background salt, the SCSD theory recovers the pure SC regime.

\begin{figure*}
\includegraphics[width=1.\linewidth]{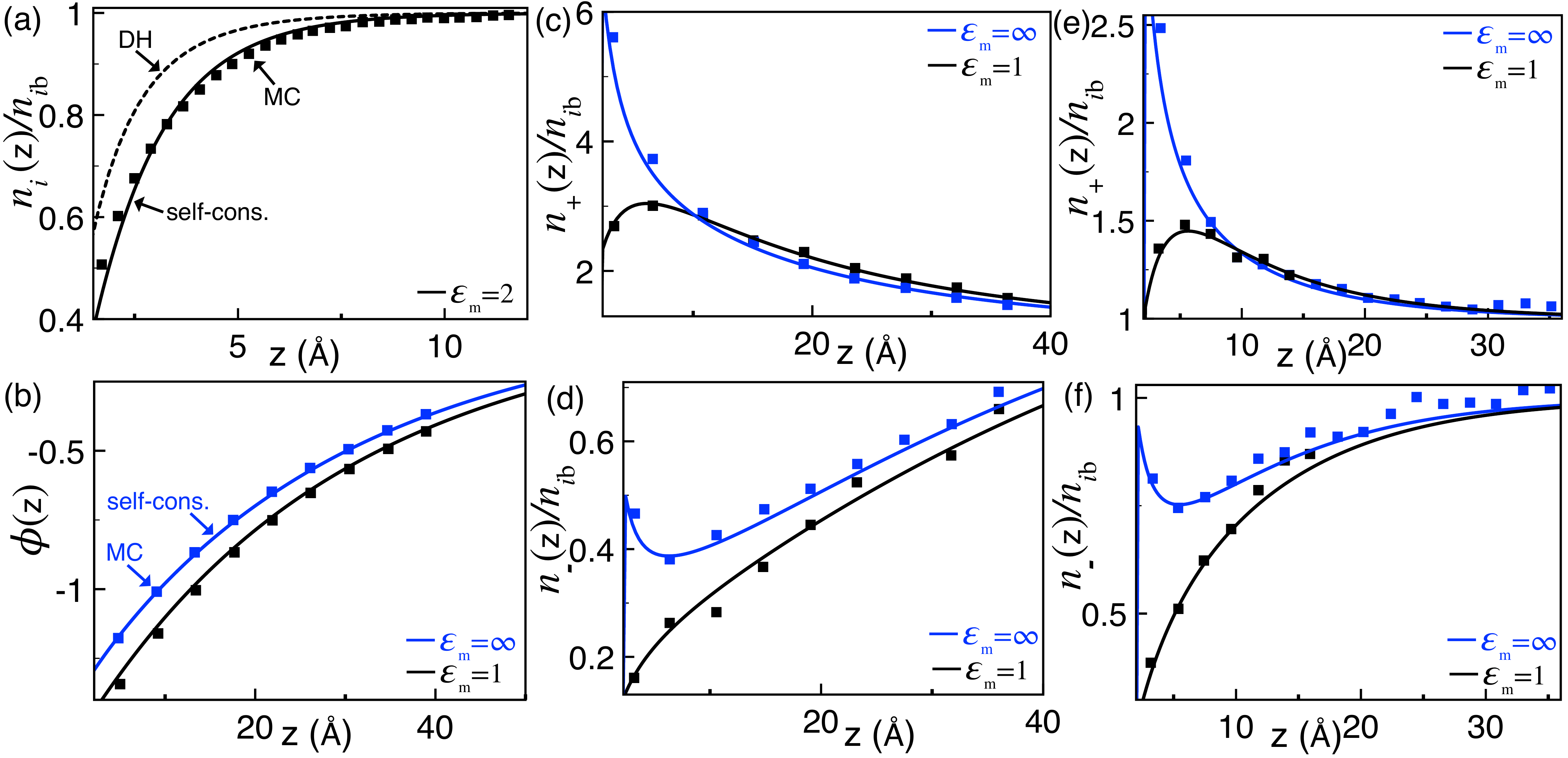}
\caption{(Color online) Ion density and potential profiles for a symmetric monovalent electrolyte at insulator ($\e_{\rm m}=1$ or $2$) and metallic membranes ($\e_{\rm m}=\infty$). (a) Dimensionless ion density at a neutral membrane surface ($\sigma\s=0$) for the bulk salt concentration $n_{i\rm b}=0.5$ M and a closest distance approach $a=1.5$ {\AA}. (b) Electrostatic potential profile, and (c) counterion and (d) coion densities at the salt concentration $n_{i\rm b}=0.01$ M, membrane charge density  $\sigma\s=0.055$ ${\rm e}/{\rm nm}^2$, and steric size $a=2.125$ {\AA}. The plots (e)-(f) are similar to the plots (c)-(d) with a higher salt concentration $n_{i\rm b}=0.1$ M and membrane charge $\sigma\s=0.0775$ ${\rm e}/{\rm nm}^2$. The symbols in (a) are the MC data of Ref.~\cite{MCneut} and the symbols in (b)-(f) are from Ref.~\cite{MCch}. The dashed curve in (a) is the DH result, and the solid curves in all plots are from the pure salt limit $n_{\rm cb}=0$ of Eqs.~(\ref{SD15II})-(\ref{SD16II}).}
\label{fig2}
\end{figure*}

\section{Results}
\label{pars}

In this section, we carry out a detailed characterization of the strongly coupled multivalent counterion effects on the monovalent salt affinity of nanopores. First, in Sec.~\ref{puresalt}, we compare with MC simulation results the predictions of Eqs.~(\ref{SD15II})-(\ref{SD16II}) on the density of purely monovalent electrolytes in contact with anionic insulator and metallic membranes. Then, in Sec.~\ref{multi}, we scrutinize the alteration of the monovalent ion partition in charged nanoslits by added multivalent counterions. Our results will be obtained from the perturbative expansion of the SCSD Eqs.~(\ref{SD15II})-(\ref{SD16II}) in terms of the multivalent counterion concentration $n_{\rm cb}$ and the recursive solution of these equations. The details of this solution scheme explained in Appendix~\ref{ap2} will not be reported here.

\subsection{Interfacial ion partition in monovalent salt solutions}
\label{puresalt}

\begin{figure*}
\includegraphics[width=1.\linewidth]{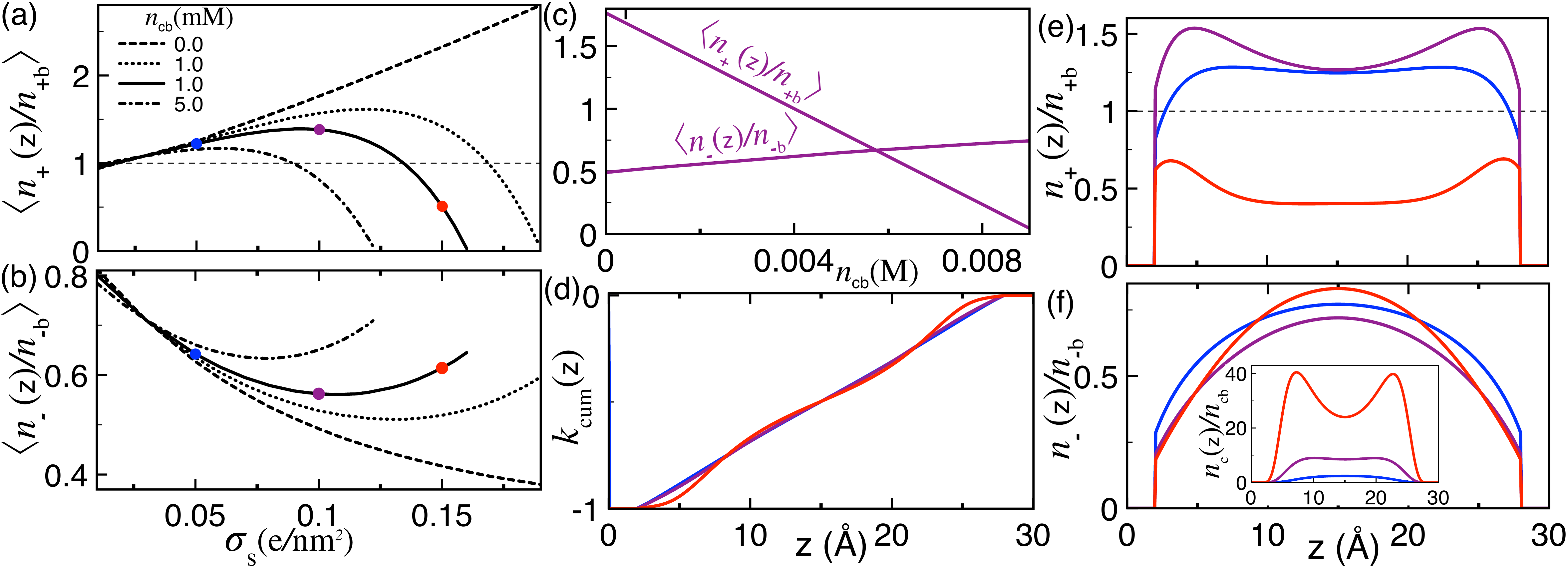}
\caption{(Color online) Pore-averaged monovalent counterion and coion densities of Eq.~(\ref{ad}) against (a)-(b) the membrane charge strength and (c) the multivalent counterion concentration. (d) Normalized cumulative charge density profile~(\ref{cum}). (e) Monovalent counterion, (f) coion (main plot) and multivalent cation density profile (inset). In (c)-(f), the membrane charge density value of each curve corresponds to the value of the circles with the same color in (a) and (b).  In all plots, the membrane permittivity is $\e_{\rm m}=1$, the bulk salt concentration $n_{+\rm b}=0.1$ M, and the pore size $d=30$ {\AA}.}
\label{fig3}
\end{figure*}
In Fig.~\ref{fig2}, we compare with MC simulations the prediction of the SCSD Eqs.~(\ref{SD15II})-(\ref{SD16II})  for the partition of a symmetric monovalent solution in contact with a dielectric plane located at $z=0$.  In the absence of multivalent ions, i.e. for $n_{\rm cb}=n\ce(\br)=0$ and $n_{-\rm b}=n_{+\rm b}$, Eqs.~(\ref{SD15II}) and~(\ref{SD16II}) tend to the WC variational equations of Ref.~\cite{NetzSC}.  We also note that in Ref.~\cite{Buyuk2012}, the MC data in Fig.~\ref{fig2}(b)-(f) for insulating membranes ($\e_{\rm m}=1$) have been used to test the truncated solution of these equations. Here, this comparison is extended to the exact numerical solution of these equations explained in Appendix~\ref{sp} as well as the case of conducting membranes with attractive image-charge interactions ($\e_{\rm m}=\infty$).

Fig.~\ref{fig2}(a) illustrates the ion partition at a neutral membrane surface of dielectric permittivity $\e_{\rm m}=2\ll\e_{\rm w}$ for the bulk salt concentration $n_{i\rm b}=0.5$ M. The comparison of the DH prediction and the MC data of Ref.~\cite{MCneut} shows that the ionic depletion originating  from image-charge interactions is underestimated by the DH result. As previously shown in Ref.~\cite{Buyuk2012} by comparison with different MC data at lower bulk salt concentrations, the inaccuracy of the DH theory originates from the uniform interfacial charge screening assumption. This approximation neglects the self-consistently reduced screening of the image-charge interactions within the ionic depletion layer and underestimates the repulsive image-charge forces. In Fig.~\ref{fig2}(a), the comparison of the solid curve and symbols shows that despite the considerably large bulk salt concentration, the enhanced interfacial depletion effect can be taken into account by the self-consistent theory  with reasonable quantitative accuracy.

Fig.~\ref{fig2}(b) compares now the average potential profiles obtained from the self-consistent formalism with the MC simulations of Ref.~\cite{MCch} at the weak membrane charge density $\sigma\s=0.055$ ${\rm e}/{\rm nm}^2$ and two different membrane permittivities. At the insulator membrane with permittivity $\e_{\rm m}=1$, the interfacial ion depletion driven by repulsive image-charge interactions weakens the screening of the average electrostatic potential. In the opposite case of metallic interfaces with permittivity $\e_{\rm m}=\infty$, attractive image-charge interactions enhancing the interfacial ion density strengthen the screening experienced by the average potential. Consequently, the potential at the insulator membrane has a larger magnitude than at the metallic interface. One notes that the self-consistent theory can account for this dielectric effect with good quantitative accuracy.

For the same membrane charge density and dielectric permittivity values, we compare in Fig.~\ref{fig2}(c) and (d) the interfacial ion density profiles obtained from the theory and simulations. In these figures, the key physical features are the counterion concentration  peak at the insulator membrane and the coion density minimum at the metallic interface. Both peculiarities are caused by the opposing effect of ion-surface charge and ion-image-charge interactions. Finally, Figs~\ref{fig2}(e)-(f) illustrate similar results at the larger ion concentration $n_{i\rm b}=0.1$ M and charge density $\sigma\s=0.0775$ ${\rm e}/{\rm nm}^2$ corresponding to a higher electrostatic coupling strength. Again, the comparison of the theoretical curves and simulation results shows that within the fluctuations of the MC data, the self-consistent theory can reproduce with good quantitative accuracy the cooperative effect of the surface polarization forces and the direct coupling between the salt ions and the fixed membrane charges. 

\subsection{Effect of multivalent counterions on the monovalent ion partition}
\label{multi}

Here, we reconsider the weak membrane charge regime of Sec.~\ref{puresalt} and investigate the alteration of the interfacial monovalent ion partition by the addition of dilute multivalent counterions. From now on, we set the steric ion size to $a=2$ {\AA} and the counterion valency to $q_{\rm c}=4$.

\subsubsection{Multivalent cation-driven coion attraction and counterion exclusion mechanism}

In Figs.~\ref{fig3}(a) and (b), we plotted the pore-averaged monovalent ion densities
\be
\label{ad}
r_i\equiv\lan \frac{n_i(z)}{n_{i\rm b}}\ran=\frac{1}{d-2a}\int_a^{d-a}\mathrm{d}z \frac{n_i(z)}{n_{i\rm b}}
\ee
versus the membrane charge density at different multivalent counterion concentrations. As expected from MF-level electrostatics, in purely monovalent solutions (dashed curves  with $n_{\rm cb}=0$), the rise of the membrane charge density amplifies monotonically the counterion attraction and the coion exclusion, i.e. $\sigma_{\rm s}\uparrow r_+\uparrow r_-\downarrow$. However, in the presence of multivalent counterions, this MF behavior is reversed beyond a characteristic surface charge density $\sigma_{\rm s}=\sigma^*_{\rm s}$ where the increment of the negative membrane charge brings further negative ions into the pore and excludes positive ions from the pore into the reservoir, i.e.  $\sigma_{\rm s}\uparrow r_+\downarrow r_-\uparrow$.  

The like-charge coion attraction and the opposite-charge monovalent counterion repulsion by the charged membrane originate from the direct salt-multivalent counterion (S-C) interactions embodied in the integral term of Eq.~(\ref{den2}). As illustrated in Fig.~\ref{fig3}(c), the sign behind this term indicates that the anion density rises and the monovalent cation density drops linearly with the amount of added multivalent counterions. In Figs.~\ref{fig3}(e) and (f), the corresponding effect is displayed with further detail in terms of the local ion partition. The density curves correspond to the dots of the same color in Figs.~\ref{fig3}(a) and (b). In the WC regime where the direct salt-membrane (S-M) interactions govern the system, the increase of the membrane charge strength from $\sigma_{\rm s}=0.05$ $e/{\rm nm}^2$ (blue curves) to $0.1$ $e/{\rm nm}^2$ (purple curves) enhances the monovalent counterion excess and the coion exclusion. Then, the further rise of the membrane charge density from $\sigma_{\rm s}=0.1$ $e/{\rm nm}^2$ (purple curves) to $0.15$ $e/{\rm nm}^2$ (red curves)  amplifies the multivalent counterion density by several factors (inset). Consequently, the S-C interactions take over the S-M coupling, attracting further coions into the mid-pore area and resulting in the overall exclusion of the counterions from the slit pore (main plots).  In Figs.~\ref{fig3}(a), one also sees that this competition between the S-M and S-C interactions leads to the drop of the critical membrane charge density with the multivalent counterion concentration, i.e. $n_{\rm cb}\uparrow\sigma^*_{\rm s}\downarrow$.

At this point, the question arises as to whether the coion attraction and counterion repulsion by the charged membrane is causally related with the membrane charge inversion phenomenon. To shed light on this point, in Fig.~\ref{fig3}(d), we plotted the rescaled cumulative charge density profile defined as
\be
\label{cum}
k_{\rm cum}(z)\equiv\frac{1}{2\sigma\s}\sum_{i=\pm,{\rm c}}q_i\int_0^d\mathrm{d}z\;n_i(z)-1.
\ee
As one moves from the bottom to the top membrane, the cumulative charge density rises monotonically from $-1$ to $0$ without turning to positive. Thus, in the characteristic membrane charge regime where the coion attraction and monovalent counterion exclusion take place (red curve), the membrane charge inversion is absent. This indicates the lack of direct correlation between these two effects. 

We finally emphasize that in Fig.~\ref{fig3}(a), the characteristic surface charge density where the monovalent counterion density cancels and turns to negative corresponds to the upper validity limit of the dilute multivalent counterion approximation introduced in the derivation of the SCSD  Eqs.~(\ref{SD15II})-(\ref{SD16II}). However, the quantitative validity of this approximation is expected to break down before this upper limit is reached.

\subsubsection{Effect of monovalent salt and pore confinement}

\begin{figure}
\includegraphics[width=1.\linewidth]{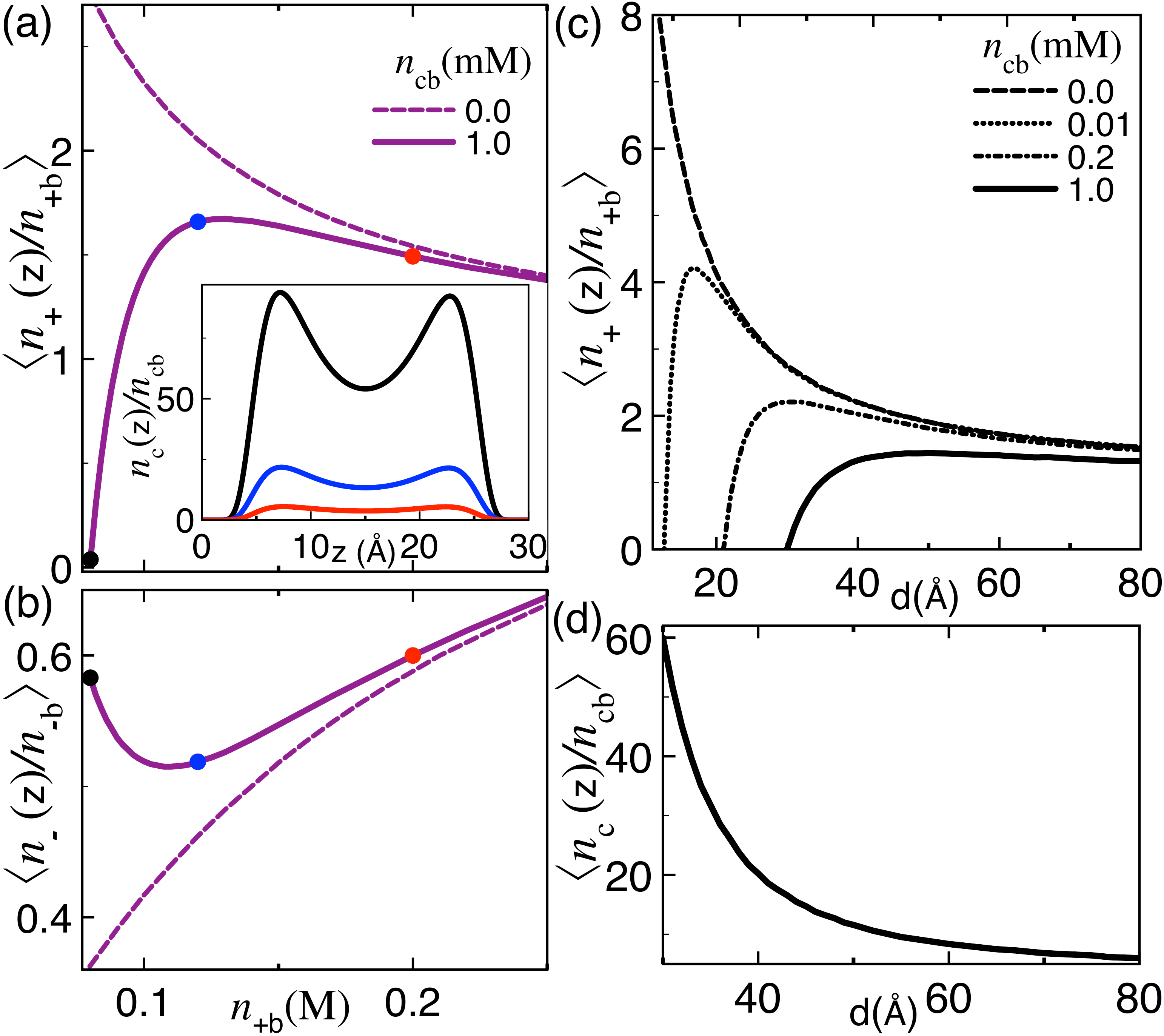}
\caption{(Color online) Pore-averaged monovalent counterion and coion densities (a)-(b) against the bulk salt concentration at the pore size $d=30$ {\AA} and (c)-(d) against the pore size at the salt concentration $n_{+\rm b}=0.08$ M. The inset in (a) illustrates the local multivalent counterion partition in the pore. The membrane charge is $\sigma_{\rm s}=0.15$ $e/{\rm nm}^2$ in all figures. The remaining parameters are the same as in Fig.~\ref{fig3}.}
\label{fig4}
\end{figure}
\begin{figure*}
\includegraphics[width=1.\linewidth]{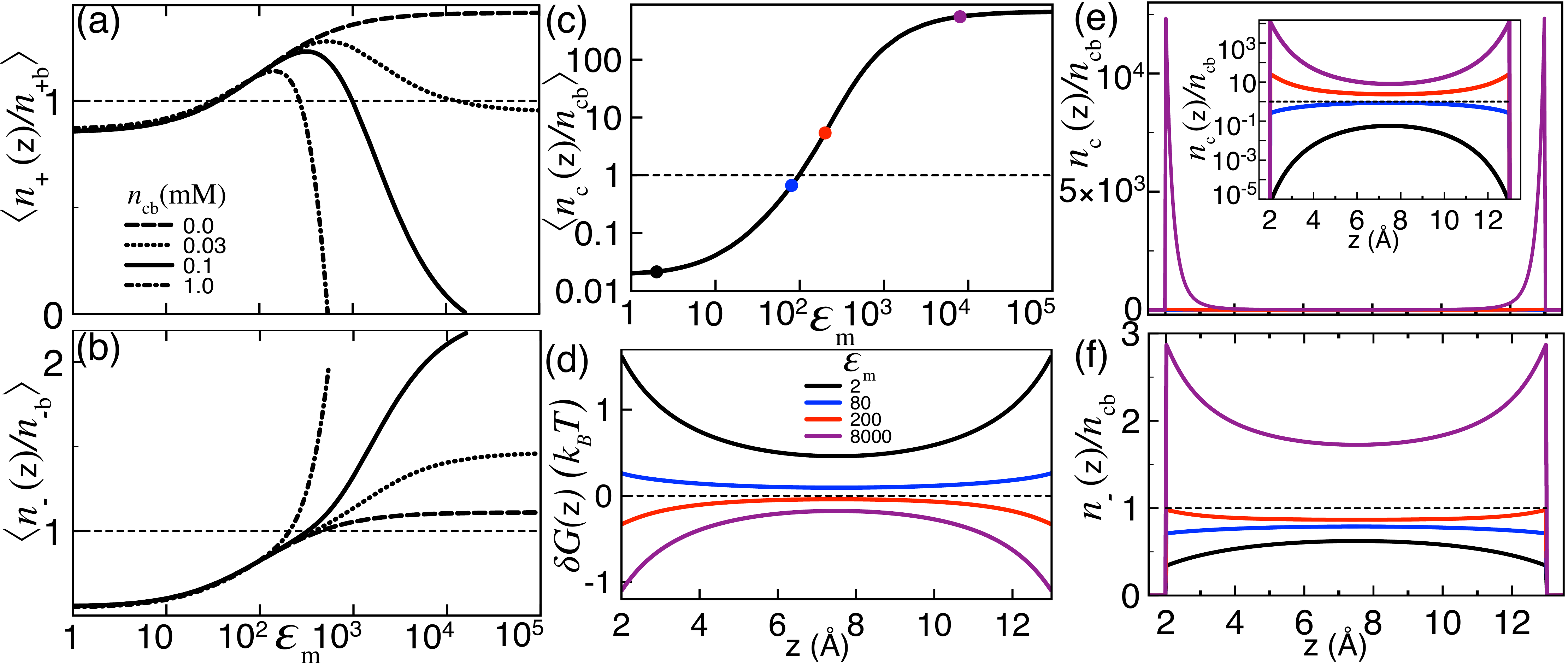}
\caption{(Color online) Pore-averaged (a) monovalent counterion and (b) coion, and (c) multivalent counterion density against the membrane permittivity $\e_{\rm m}$. (d) Ionic self-energy, (e) multivalent counterion density, and (f) coion density profiles at the membrane permittivities displayed in the legend of (d). The membrane charge is $\sigma_{\rm m}=0.01$ $e/{\rm nm}^2$, the salt concentration $n_{+ \rm b}=0.1$ M, and the pore size $d=15$ {\AA}. In (c)-(f), the bulk counterion density  is $n_{\rm cb}=0.1$ mM.}
\label{fig5}
\end{figure*}

We probe now the influence of the bulk salt concentration on the counterion-induced correlation effects. Figs.~\ref{fig4}(a) and (b) display the salt dependence of the pore-averaged monovalent ion densities (main plots) and the local multivalent counterion density (inset). In a purely monovalent electrolyte governed by S-M interactions (dashed curves), salt addition into the reservoir screens these interactions and weakens monotonously the WC-level monovalent counterion attraction and coion repulsion effects, $n_{+\rm b}\uparrow r_+\downarrow r_-\uparrow$. However, in the presence of multivalent counterions (solid curves), monovalent ion densities exhibit a non-monotonic dependence on the bulk salt concentration. Namely, rising the salt concentration from the value $n_{+\rm b}=0.08$ M associated with the counterion exclusion regime (black curves and dots) to $n_{+\rm b}=0.12$ M (blue curves and dots), the screening of the membrane surface charge lowers significantly the multivalent counterion density in the pore, $n_{+\rm b}\uparrow n\ce(z)\downarrow$. This weakens the strength of the S-C interactions and attenuates the monovalent counterion exclusion and coin attraction driven by these interactions, i.e. $n_{+\rm b}\uparrow r_+\uparrow r_-\downarrow$.  Beyond the salt concentration $n_{+\rm b}\approx0.12$ M where the multivalent counterion density in the pore is considerably reduced (see the red curve in the inset), the S-M interactions entirely take over the S-C coupling. This drives the system to the WC regime of pure monovalent solutions where $n_{+\rm b}\uparrow r_+\downarrow r_-\uparrow$.

Figs.~(\ref{fig4})(c) and (d) show that the pore confinement has a similar effect on the monovalent salt partition. In the absence of multivalent charges ($n_{\rm cb}=0$) where the salt affinity of the pore is set by the WC S-M interactions, the decrease of the pore size amplifies the average electrostatic potential and enhances monotonically the monovalent counterion density, $d\downarrow r_+\uparrow$. However, in solutions including multivalent counterions ($n_{\rm cb}>0$), this MF behavior is bounded by a lower pore size $d=d^*$ where the average counterion density $r_+$ reaches a maximum. Indeed, below this pore size, the high electrostatic potential leads to a pronounced multivalent counterion adsorption, $r\ce\gg1$. Consequently, the repulsion of the monovalent counterions by the multivalent counterions takes over their attraction by the membrane charges. This results in the drop of the monovalent counterion density with the pore size, i.e. $d\downarrow r\ce\uparrow r_+\downarrow$.  Finally, Fig.~(\ref{fig4})(c) shows that the corresponding competition between the S-M and S-C interactions leads to the rise of the critical pore size with the bulk counterion concentration, i.e. $n_{\rm cb}\uparrow d^*\uparrow$. The effect of the surface polarization forces on this competition is investigated in Sec.~\ref{pol}.

\subsubsection{Membrane permittivity and surface polarization forces}
\label{pol}

Membrane engineering techniques based on the insertion of graphene nanoribbons into a host matrix allows to increase the dielectric permittivity of the substrate from the characteristic range of biological and synthetic membranes $\e_{\rm m}\sim1$ up to the value of $\e_{\rm m}\approx8000$ comparable with the metallic interface limit~\cite{Dang,Dimiev}. Motivated by this point, we consider a weak pore surface charge $\sigma\s=0.01$ $e/{\rm nm}^2$ and characterize the influence of the dielectric contrast between the membrane and the solvent on the multivalent counterion-driven correlations.  

Figs.~\ref{fig5}(a)-(c) illustrate the pore-averaged mono- and multivalent ion densities $r_{\pm,\rm c}$ versus the membrane permittivity $\e_{\rm m}$. Fig.~\ref{fig5}(d) displays in turn the ionic self-energy embodying the image-charge interactions and obtained from the equal point correlation function renormalized by its bulk limit, $\delta G(z)=\left[G(\br,\br')-G\B(\br-\br')\right]_{\br'\to\br}$. One sees that the rise of the membrane permittivity switches the self-energy from repulsive to attractive, $\e_{\rm m}\uparrow\delta G(z)\downarrow$. In a monovalent electrolyte (dashed curves at $n_{\rm cb}=0$), this rises weakly the pore-averaged ion densities, i.e. $\e_{\rm m}\uparrow r_\pm\uparrow$. However, the curves for $n_{\rm cb}>0$ shows that this trend is radically modified by the presence of multivalent counterions. Indeed, Fig.~\ref{fig5}(c) indicates that upon addition of bulk multivalent counterions, the attractive polarization forces at $\e_{\rm m}>\e_{\rm w}$ lead to the adsorption of these charges from the reservoir into the pore, i.e. $\e_{\rm m}\uparrow r_{\rm c}\uparrow$. Consequently, beyond a characteristic permittivity $\e_{\rm m}^*\approx 200$ where $r\ce\gg1$, the interaction of the monovalent ions with these adsorbed multivalent cations takes over their interaction with their image charges. As a result, the rise of the membrane permittivity beyond $\e_{\rm m}^*$ strengthens the coion adsorption and triggers the monovalent counterion exclusion, i.e. $\e_{\rm m}\uparrow  r_-\uparrow r_+\downarrow$. Figs.~\ref{fig5}(a)-(b) also shows that the multivalent counterion abundance in the reservoir reduces the critical permittivity, i.e. $n_{\rm cb}\uparrow\e_{\rm m}^*\downarrow$.

The comparison of Figs.~\ref{fig5}(a)-(b) with Figs.~\ref{fig3}(a)-(b) indicates that the coion attraction and counterion repulsion effects mediated by the image-charge-driven multivalent counterions occur more accutely than their counterpart induced by the membrane-charge-driven counterion adsorption. One indeed notes that only the former gives rise to a net coion excess  in the pore ($r_->1$). To shed light on this peculiarity, in Figs.~\ref{fig5}(e) and (f), we plotted the local multivalent counterion and coion densities in the nanoslit.  One sees that the layer of counterions attracted by the image-charge forces exhibits a more localized interfacial structure and a much higher peak than in the inset of Fig.~\ref{fig3}(f). This stems from the fact that the attractive image-charge energy $q\ce^2\delta G(z)\sim -q\ce^2 e^{-2\kappa\s z}$ scaling quadratically with the counterion valency has a shorter range  but stronger amplitude than the counterion-surface charge interactions characterized by the asymptotic large distance behavior $q\ce\phi_0(z)\sim -q\ce e^{-\kappa\s z}$~\cite{Buyuk2012}. The inset of Fig.~\ref{fig5}(e) shows that this gives rise to an interfacial counterion density exceeding the density in the mid-pore region by three orders of magnitude (purple curve). Fig.~\ref{fig5}(f) reveals that unlike the weak coion adsorption observed in Fig.~\ref{fig3}(f), these interfacial counterion layers acting as effective positive surface charges lead to a substantial coion adsorption through the entire pore region.

\section{Conclusions}

Experiments on charged macromolecules often involve composite electrolytes characterized by the coexistence of ionic species with different valencies and electrostatic coupling strength~\cite{Heyden2006,Qiu2015,exp2,exp3}. The interpretation of these experiments thus requires the formulation of electrostatic theories able to handle self-consistently multiple interaction strengths between the mobile ions and the macromolecular surfaces. Motivated by this need, in this article, we derived the SD equations for electrolyte mixtures composed of mono- and multivalent ions confined to charged nanopores. The WC interactions between the monovalent ions and the nanopore charges were treated within a gaussian approximation assuming moderate fluctuations of the electrostatic potential around the MF potential. The multivalent counterions strongly coupled to the slit charges were in turn treated within the SC approximation based on a low fugacity expansion. 

As the fugacity expansion incorporates the mono- and multivalent ion interactions via the integral of the multivalent counterion density over the entire pore, the SCSD Eqs.~(\ref{SD15II}) and~(\ref{SD16II}) emerged in a non-local form. In order to solve these integro-differential equations, we developed a general recursive scheme that can be adapted to all geometries relevant to experiments and simulations. In the present work, we explored the predictions of the SCSD theory in the slit pore geometry.

First, by comparison with MC simulations, we verified the accuracy of the SCSD equations in predicting the partition of purely monovalent solutions in contact with weakly charged insulator and conductor membrane surfaces. Then, in the same membrane charge regime, we focused on the confined slit geometry and characterized the alteration of the monovalent ion distribution by the presence of multivalent counterions in the solution. We showed that added multivalent cations systematically enhance the anion density and reduce the monovalent cation density in the pore. The overall monovalent salt affinity of the pore is characterized by the hierarchy between the S-C interactions responsible for this effect, and the S-M interactions driving the WC-level monovalent counterion attraction and coion repulsion.  

Beyond a characteristic membrane charge strength where the multivalent counterion adsorption into the pore becomes significant, the S-C interactions take over the S-M coupling.  As a result, a stronger negative membrane charge brings additional negative ions into the pore and drives the positive charges from the pore into the reservoir. This like-charge coion adsorption and opposite-charge counterion exclusion effect is the key physical prediction of our work. Via the alteration of the multivalent counterion adsorption setting the balance between the S-M and S-C interactions, this effect is suppressed by monovalent salt addition into the reservoir but strengthened by the pore confinement. Moreover, the effect becomes particularly strong with attractive polarization forces emerging in the {\it giant dielectric permittivity} regime of engineered membranes~\cite{Dang}. 

Our results may provide guiding information for future simulations and experiments where the monovalent coion adsorption and counterion exclusion effect can be easily probed. We also emphasize that considering the sizable influence of monovalent salt on electrostatic interactions, this newly predicted effect may have important repercussions on various  electrostatically driven phenomena ranging from macromolecular interactions in gene therapeutic applications to ion and polymer transport in nanofluidics and nanopore-based sequencing approaches. We plan to explore the applications of our formalism to these systems in future works. 
\\

{\bf Acknowledgement.} I acknowledge many stimulating discussions with R. Podgornik on charge correlations and SC electrostatics in background salt. I would also like to thank the Beijing Computational Science Research Center and the University of Chinese Academy of Sciences for their hospitality during an academic visit in 2019 where some part of this work has been accomplished.

\smallskip
\appendix
\section{Derivation of the SD equations for a general functional integral}
\label{ap1}

We review here the derivation of the SD equalities associated with a general functional integral of the form~(\ref{zg2})~\cite{justin}. To this end, we introduce first the integral
\be
\label{a1}
I=\int \mathcal{D}\phi\mathcal{D}\psi\;e^{-H[\phi,\psi]}F[\phi,\psi]
\ee
where $F[\phi,\psi]$ and $H[\phi,\psi]$ are general functionals of the potentials $\phi(\br)$ and $\psi(\br)$. Computing now the variation of the integral~(\ref{a1}) under the infinitesimal shift of the electrostatic potential $\phi(\br)\to\phi(\br)+\delta\phi(\br)$, one gets
\bea
\label{a2}
\delta I&=&\int \mathcal{D}\phi\mathcal{D}\psi\;\exp\left\{-H[\phi,\psi]-\int\mathrm{d}\br\frac{\delta H[\phi,\psi]}{\delta \phi(\br)}\delta \phi(\br)\right\}\nonumber\\
&&\hspace{1cm}\times\left\{F[\phi,\psi]-\int\mathrm{d}\br\frac{\delta F[\phi,\psi]}{\delta \phi(\br)}\delta \phi(\br)\right\}\nonumber\\ 
&&-\int \mathcal{D}\phi\mathcal{D}\psi\;e^{-H[\phi,\psi]}F[\phi,\psi].
\eea
At the next step, we expand the r.h.s. of Eq.~(\ref{a2}) at the linear order in $\delta\phi(\br)$ to obtain 
\bea
\label{a3}
\delta I&=&\int\mathrm{d}\br\delta \phi(\br)\int \mathcal{D}\phi\mathcal{D}\psi\;e^{-H[\phi,\psi]}\\
&&\hspace{2cm}\times\left\{\frac{\delta F[\phi,\psi]}{\delta \phi(\br)}-F[\phi,\psi]\frac{\delta H[\phi,\psi]}{\delta \phi(\br)}\right\}.\nonumber
\eea
Then, we note that the shift $\delta\phi(\br)$ can be absorbed into the redefinition of the functional integral measure in Eq.~(\ref{a1}). This means that the integral~(\ref{a1}) is left invariant by the shift $\delta\phi(\br)$, i.e. $\delta I=0$. Consequently, dividing Eq.~(\ref{a3}) by the partition function~(\ref{zg2}) and setting the result to zero, one obtains the identity
\be\label{a4}
\lan \frac{\delta F[\phi,\psi]}{\delta \phi(\br)}\ran=\lan F[\phi,\psi]\frac{\delta H[\phi,\psi]}{\delta \phi(\br)}\ran
\ee
where the bracket symbol designates the field average defined by Eq.~(\ref{av1}). If one now sets in Eq.~(\ref{a4}) $F[\phi]=1$ and $F[\phi]=\phi(\br')$, one obtains respectively the SD Eqs.~(\ref{SD1}) and~(\ref{SD2}) of the main text.

\section{Perturbative solution of the SCSD equations}
\label{ap2}

We explain here the solution of the SCSD Eqs.~(\ref{SD15II}) and~(\ref{SD16II}) via their expansion in terms of the multivalent counterion concentration $n_{\rm cb}$. First in Sec.~\ref{apII}, the SCSD Eqs.~(\ref{SD15II})-(\ref{SD16II})  are solved in a bulk reservoir. Then, this solution is used in Sec.~\ref{apIII} in order to reexpress these equations by replacing the ionic fugacities $\Lambda_i$ with the bulk ion concentrations $n_i$. In Secs.~\ref{apIV} and~\ref{anel}, the resulting equations and local ion density functions are expanded at the linear order in the multivalent ion concentration. Finally, in Sec.~\ref{nums}, we introduce an iterative scheme for the solution of the expanded SCSD equations.

\subsection{Solving the SCSD Eqs.~(\ref{SD15II})-(\ref{SD16II}) in bulk liquids}
\label{apII}

Our formalism was derived in the grand canonical ensemble where the mobile ions located in the vicinity of charged macromolecules or confined to a nanopore are in chemical equilibrium with the bulk reservoir. Thus, in order to relate the ionic fugacities $\Lambda_i$ in Eqs.~(\ref{denc})-(\ref{denibare}) and~(\ref{den2}) fixed by this equilibrium to the bulk ion concentrations $n_{i\rm b}$, we will solve here the SCSD Eqs.~(\ref{SD15II}) and~(\ref{SD16II}) in the bulk reservoir.

In a bulk electrolyte characterized by a vanishing macromolecular charge $\sigma(\br)=0$ and steric potential $V_i(\br)=0$, and uniform dielectric permittivity $\e(\br)=\e_{\rm w}$, the average potential vanishes, and the medium is governed by the spherical symmetry, i.e.
\bea
\label{b1}
\phi_0(\br)&=&0,\\
\label{b2}
G(\br,\br')&=&G\B(\br-\br')=\int\frac{\mathrm{d}\bq}{(2\pi)^3}\tG(q)e^{i\bq\cdot\left(\br-\br'\right)}.
\eea
Eq.~(\ref{b2}) corresponds to the Fourier transform of the bulk propagator. Furthermore, in the same bulk region, the ionic distribution functions~(\ref{denc})-(\ref{denibare}) and~(\ref{den2}) become 
\bea
\label{b3}
\rho_{\pm\rm b}&=&\Lambda_\pm e^{-\frac{1}{2}\left[G\B(0)-v\ce(0)\right]},\\
\label{b4}
n_{\rm cb}&=&\Lambda\ce e^{-\frac{q^2\ce}{2}\left[G\B(0)-v\ce(0)\right]},\\
\label{b5}
 n_{\pm\rm b}&=&\rho_{\pm\rm b}\left[1\mp q\ce n_{\rm cb}\tG\B(0)\right].
\eea

Injecting Eqs.~(\ref{b1}) and~(\ref{b2}) into the SCSD Eqs.~(\ref{SD15II}) and~(\ref{SD16II}), one gets
\bea
\label{b6}
&&\rho_{+\rm b}=\rho_{-\rm b},\\
\label{b7}
&&q\ce n_{\rm cb}\left(\rho_{+\rm b}-\rho_{-\rm b}\right)\tG\B(0)\tG\B(q)=\frac{\kappa\s^2+q^2}{4\pi\ell_{\rm B}}\tG\B(q)-1,\nonumber\\
\eea
where we defined the DH screening parameter associated with salt,
\be
\label{b8}
\kappa^2\s=4\pi\ell_{\rm B}\left(\rho_{+\rm b}+\rho_{-\rm b}\right).
\ee
Combining Eqs.~(\ref{b6}) and~(\ref{b7}), one obtains
\be
\label{b10}
\tG\B(q)=\frac{4\pi\ell_{\rm B}}{\kappa\s^2+q^2}.
\ee
Moreover, Eqs.~(\ref{b5}) and~(\ref{b6}) yield $\rho_{+\rm b}+\rho_{-\rm b}= n_{+\rm b}+ n_{-\rm b}$. The latter identity allows to express the salt screening parameter~(\ref{b8}) in terms of the physical bulk salt concentrations $n_{\pm\rm b}$ as
\be
\label{b12}
\kappa^2\s=4\pi\ell_{\rm B}\left( n_{+\rm b}+ n_{-\rm b}\right).
\ee
Plugging Eq.~(\ref{b10}) into Eq.~(\ref{b2}), the bulk Green's function follows in the form of the Debye potential as
\be
\label{b13}
G\B(\br-\br')=\frac{\ell_{\rm B}}{|\br-\br'|}e^{-\kappa\s|\br-\br'|}.
\ee

If one now inverts Eqs.~(\ref{b4}) and~(\ref{b5}), the charge fugacities follow at the order $O\left(n_{\rm cb}\right)$ in the form
\bea
\label{b14}
\Lambda\ce&=& n_{\rm cb}e^{\frac{q^2\ce}{2}\left[G\B(0)-v\ce(0)\right]},\\
\label{b15}
\Lambda_\pm&=& n_{\pm\rm b}e^{\frac{1}{2}\left[G\B(0)-v\ce(0)\right]}\left[1\pm q\ce n_{\rm cb}\tG\B(0)\right]\\
&=& n_{\pm\rm b}e^{\frac{1}{2}\left[G\B(0)-v\ce(0)\right]}\left[1\pm q\ce\int\mathrm{d}\brc n_{\rm cb}G\B(\br-\brc)\right].\nonumber
\eea
Finally, combining Eqs.~(\ref{b5}) and~(\ref{b6}), and using Eq.~(\ref{b10}), the electroneutrality condition in the bulk reservoir follows as
\be\label{b16}
 n_{+\rm b}+q\ce n_{\rm cb}- n_{-\rm b}=0.
\ee
This shows that the SCSD Eqs.~(\ref{SD15II})-(\ref{SD16II}) assure consistently the bulk electroneutrality. The equalities~(\ref{b12})-(\ref{b16}) derived here will be used below for the solution of the SCSD Eqs.~(\ref{SD15II})-(\ref{SD16II}) in inhomogeneous medium. 

\subsection{Replacing the ionic fugacities by the reservoir concentrations}
\label{apIII}

We express now the SCSD Eqs.~(\ref{SD15II}) and~(\ref{SD16II}) in terms of the bulk ion concentrations. First, injecting Eqs.~(\ref{b14}) and~(\ref{b15}) into the density functions in Eqs.~(\ref{denc})-(\ref{denibare}) and~(\ref{den2}), one gets at the order $O(n_{\rm cb})$
\bea
\label{d1}
n\ce(\br)&=&n_{\rm cb}k\ce(\br),\\
\label{d2}
n_\pm(\br)&=&n_{\pm\rm b}k_\pm(\br)\\
&&\times\left\{1\mp q\ce n_{\rm cb}\int\mathrm{d}\br\ce\left[k\ce(\brc)G(\br,\brc)-G_{\rm b}(\br-\brc)\right]\right\}\nonumber,\\
\label{d3}
\rho_\pm(\br)&=&n_{\pm\rm b}k_\pm(\br)\left[1\pm q\ce n_{\rm cb}\int\mathrm{d}\br\;G_{\rm b}(\br-\brc)\right],
\eea
where we defined the ionic partition function
\be
k_i(\br)=e^{-V_i(\br)-q_i\phi_0(\br)-\frac{q_i^2}{2}\delta G(\br)}
\ee
for $i=\{\pm,{\rm c}\}$, with the ionic self-energy
\be\label{se}
\delta G(\br)\equiv\lim_{\br'\to\br}\left[G(\br,\br')-G\B(\br-\br')\right].
\ee
Substituting Eqs.~(\ref{d1})-(\ref{d3}) into the SCSD Eqs.~(\ref{SD15II})-(\ref{SD16II}), the latter finally take the following form,
\begin{widetext}
\bea
\label{SD21}
&&\frac{k_{\rm B}T}{e^2}\nabla_\br\cdot\e(\br)\nabla_\br\phi_0(\br)+\sum_{i=\pm}q_in_{i\rm b}k_i(\br)+\sigma(\br)=-q\ce n_{\rm cb}\sum_{i=\pm}q_i^2n_{i\rm b}k_i(\br)\int\mathrm{d}\brc\;G_{\rm b}(\br-\brc),\\
\label{SD22}
&&\left\{\frac{k_{\rm B}T}{e^2}\nabla_\br\cdot\e(\br)\nabla_\br-\sum_{i=\pm}q_i^2n_{i\rm b}k_i(\br)\right\}G(\br,\br')+\delta(\br-\br')\nonumber\\
&&=-q\ce n_{\rm cb}\sum_{i=\pm}q_i^3n_{i\rm b}k_i(\br)\int\mathrm{d}\br\ce\left[k\ce(\brc)G(\br,\brc)-G_{\rm b}(\br-\brc)\right]G(\br,\br').
\eea
\end{widetext}

\subsection{Expansion of the SCSD Eqs.~(\ref{SD21}) and~(\ref{SD22}) in terms of the counterion concentration $n_{\rm cb}$}
\label{apIV}

In this Appendix, we carry out the systematic expansion of the SCSD Eqs.~(\ref{SD21}) and~(\ref{SD22}) in terms of the multivalent counterion concentration $n_{\rm cb}$. From now on, we assume that all ionic species have access to the same configurational space, i.e. $V_i(\br)=V(\br)$. First, we express the electrostatic potential and ionic partition functions as the superposition of a monovalent salt component and a multivalent counterion contribution, i.e.
\bea
\label{sp1}
\phi_0(\br)&=&\phi\s(\br)+\phi\ce(\br),\\
\label{sp2}
G(\br,\br')&=&G\s(\br,\br')+G\ce(\br,\br'),\\
\label{sp3}
k_i(\br)&=&k_{i \rm s}(\br)+k_{i \rm c}(\br),
\eea
with the components of the ionic partition functions
\bea
\label{sp4}
k_{i \rm s}(\br)&=&e^{-V(\br)-q_i\phi\s(\br)-\frac{q_i^2}{2}\delta G\s(\br)},\\
\label{sp5}
k_{i \rm c}(\br)&=&k_{i \rm s}(\br)\left\{-q_i\phi\ce(\br)-\frac{q_i^2}{2}\delta G\ce(\br)\right\},
\eea
for $i=\{\pm,{\rm c}\}$.
Next, we insert Eqs.~(\ref{sp1})-(\ref{sp3}) into Eqs.~(\ref{SD21})-(\ref{SD22}) together with the relation $n_{-\rm b}=n_{+\rm b}+q\ce n_{\rm cb}$ that follows from Eq.~(\ref{b16}). Linearizing the result in terms of the counterion concentration $n_{\rm cb}$, one gets at the order $O(n_{\rm cb}^0)$ the WC self-consistent equations associated with the monovalent salt~\cite{Netz2003},
\bea
\label{SD23}
&&\frac{k_{\rm B}T}{e^2}\nabla_\br\e(\br)\nabla_\br\phi\s(\br)-2n_{+\rm b}k_0(\br)\sinh\left[\phi\s(\br)\right]=-\sigma(\br),\nonumber\\
&&\\
\label{SD24}
&&\left\{\frac{k_{\rm B}T}{e^2}\nabla_\br\e(\br)\nabla_\br-2n_{+\rm b}k_0(\br)\cosh\left[\phi\s(\br)\right]\right\}G\s(\br,\br')\nonumber\\
&&=-\delta(\br-\br'),
\eea
and the terms of order $O(n_{\rm cb})$ yield the electrostatic equations associated with the SC counterions,
\bea
\label{SD25}
&&\left\{\frac{k_{\rm B}T}{e^2}\nabla_\br\e(\br)\nabla_\br-2n_{+\rm b}k_0(\br)\cosh\left[\phi\s(\br)\right]\right\}\phi\ce(\br)\nonumber\\
&&=\left\{q\ce n_{\rm cb}-n_{+\rm b}\delta G\ce(\br)\right\}k_0(\br)\sinh\left[\phi\s(\br)\right],\\
\label{SD26}
&&\left\{\frac{k_{\rm B}T}{e^2}\nabla_\br\e(\br)\nabla_\br-2n_{+\rm b}k_0(\br)\cosh\left[\phi\s(\br)\right]\right\}G\ce(\br,\br')\nonumber\\
&&=S(\br)G\s(\br,\br').
\eea

In Eqs.~(\ref{SD23})-(\ref{SD26}), we defined the auxiliary functions
\bea
\label{sp6}
k_0(\br)&=&e^{-V(\br)-\frac{1}{2}\delta G\s(\br)},\\
\label{sp6II}
\delta G\ce(\br)&=&\lim_{\br'\to\br}\left[G\ce(\br,\br')-G_{\rm cb}(\br-\br')\right],\\
\label{sp7}
S(\br)&=&2n_{+\rm b}k_0(\br)\sinh\left[\phi\s(\br)\right]\\
&&\times\left\{\phi\ce(\br)+q\ce n_{\rm cb}\int\mathrm{d}\br\ce k_{\rm cs}(\brc)G\s(\br,\brc)\right\}\nonumber\\
&&+\left\{q\ce n_{\rm cb}-n_{+\rm b}\delta G\ce(\br)\right\}k_0(\br)\cosh\left[\phi\s(\br)\right],\nonumber\\
\label{sp8}
k_{\rm cs}(\br)&=&e^{-V(\br)-q\ce\phi\s(\br)-\frac{q\ce^2}{2}\delta G\s(\br)}.
\eea
Eqs.~(\ref{sp6II})-(\ref{sp7}) contain the bulk limit of the counterion contribution to the Green's function $G_{\rm cb}(\br-\br')$. The explicit form of this bulk Green's function will be derived below. Finally, using Eq.~(\ref{SD24}) and the definition of the Green's operator
\be
\label{inv}
\int\mathrm{d}\br''G_{\rm s}^{-1}(\br,\br'')G_{\rm s}(\br'',\br')=\delta(\br-\br'),
\ee
the differential Eqs.~(\ref{SD25})-(\ref{SD26}) can be recast as integral equations more adequate for numerical treatment,
\bea
\label{SD27}
\phi\ce(\br)&=&-\int\mathrm{d}\br_1G\s(\br,\br_1)k_0(\br_1)\sinh\left[\phi\s(\br_1)\right]\nonumber\\
&&\hspace{1cm}\times\left[q\ce n_{\rm cb}-n_{+\rm b}\delta G\ce(\br_1)\right],\\
\label{SD28}
G\ce(\br,\br')&=&-\int\mathrm{d}\br_1G\s(\br,\br_1)S(\br_1)G\s(\br_1,\br').
\eea
The solution of Eqs.~(\ref{SD23})-(\ref{SD24}) and~(\ref{SD27})-(\ref{SD28}) will be explained in Appendix~\ref{nums}.

\subsection{Fulfillment of the global electroneutrality}
\label{anel}

Before explaining the solution of Eqs.~(\ref{SD23})-(\ref{SD24}) and~(\ref{SD27})-(\ref{SD28}), we will show that these equations consistently satisfy the global electroneutrality condition. To this end, we substitute Eqs.~(\ref{sp1})-(\ref{sp3}) and the relation $n_{-\rm b}=n_{+\rm b}+q\ce n_{\rm cb}$ into Eqs.~(\ref{d1}) and~(\ref{d2}). Expanding the result at the linear order in the counterion concentration $n_{\rm cb}$, the local ion density functions corresponding the approximation level of the expanded SDSC Eqs.~(\ref{SD23})-(\ref{SD24}) and~(\ref{SD27})-(\ref{SD28}) follow as
\bea
\label{sp9}
n\ce(\br)&=&n_{\rm cb}k_{\rm cs}(\br),\\
\label{sp10}
n_{+}(\br)&=&n_{+\rm b}k_{+\rm s}(\br)\\
&&\hspace{0cm}\times\left\{1-\phi\ce(\br)-\frac{1}{2}\delta G\ce(\br)\right.\nonumber\\
&&\hspace{0cm}\left.-q\ce n_{\rm cb}\int\mathrm{d}\br\ce \left[k_{\rm cs}(\brc)G\s(\br,\brc)-G_{\rm sb}(\br-\brc)\right]\nonumber\right\},\\
\label{sp11}
n_{-}(\br)&=&n_{+\rm b}k_{-\rm s}(\br)\\
&&\hspace{0cm}\times\left\{1+\phi\ce(\br)-\frac{1}{2}\delta G\ce(\br)\right.\nonumber\\
&&\hspace{0cm}\left.+q\ce n_{\rm cb}\int\mathrm{d}\br\ce \left[k_{\rm cs}(\brc)G\s(\br,\brc)-G_{\rm sb}(\br-\brc)\right]\nonumber\right\}\nonumber\\
&&+q\ce n_{\rm cb}k_{-\rm s}(\br).\nonumber
\eea
The bulk Green's function in Eqs.~(\ref{sp10})-(\ref{sp11}) can be obtained from the bulk solution of Eq.~(\ref{SD24}) as 
\be
\label{b13}
G_{\rm sb}(\br-\br')=\frac{\ell_{\rm B}}{|\br-\br'|}e^{-\kappa_0|\br-\br'|},
\ee
with the screening parameter corresponding to a symmetric salt solution with ion concentrations $n_{-\rm b}=n_{+\rm b}$,
\be\label{r2}
\kappa_0=\sqrt{8\pi\ell_Bn_{+\rm b}}.
\ee

The net mobile charge density function is given by
\be
\label{net}
Q\ce(\br)=\sum_{i=\pm,{\rm c}}q_in_i(\br).
\ee
Plugging Eqs.~(\ref{sp9})-(\ref{sp11}) into Eq.(\ref{net}), and using the differential equations~(\ref{SD23})-(\ref{SD26}) satisfied by the electrostatic potentials, after some algebra, the mobile charge density~(\ref{net}) takes the form
\bea
\label{net2}
Q\ce(\br)&=&-\sigma(\br)\\
&&-\frac{k_{\rm B}T}{e^2}\nabla_\br\cdot\e(\br)\nabla_\br\left\{\phi\s(\br)+\phi\ce(\br)\right.\nonumber\\
&&\hspace{2cm}\left.+q\ce n_{\rm cb}\int\mathrm{d}\br\ce k_{\rm cs}(\brc)G\s(\br,\brc)\right\}.\nonumber
\eea
Integrating now Eq.~(\ref{net2}) over the entire volume $V$ of the system, and using the divergence theorem, one obtains
\bea\label{net3}
&&\int_V\mathrm{d}\br\left[Q\ce(\br)+\sigma(\br)\right]\\
&&=-\frac{k_{\rm B}T}{e^2}\int_{S(V)}\mathrm{d}\mathbf{S}\cdot\e(\br)\nabla_\br\left\{\phi\s(\br)+\phi\ce(\br)\right.\nonumber\\
&&\left.\hspace{2.cm}\left.+q\ce n_{\rm cb}\int\mathrm{d}\br\ce k_{\rm cs}(\brc)G\s(\br,\brc)\right\}\right|_{\br\in S(V)},\nonumber
\eea
where $S(V)$ is the surface of the system boundary. Due to the absence of charges and finite electric field at this boundary, the r.h.s. of Eq.~(\ref{net3}) vanishes. This finally yields the global electroneutrality condition
\be
\label{net4}
\int_V\mathrm{d}\br\left[Q\ce(\br)+\sigma(\br)\right]=0
\ee
indicating that the mobile charge density exactly compensates the total macromolecular charge. In the case of the nanopore geometry, the l.h.s. of Eq.~(\ref{net4}) corresponds to the integral over the pore volume confining the mobile and fixed charges. This shows that Eqs.~(\ref{SD23})-(\ref{SD26}) assure the pore electroneutrality condition. We additionally note that the equality~(\ref{net4}) is illustrated in Fig.~\ref{fig3}(d) for the specific slit pore geometry.

\subsection{Numerical solution of Eqs.~(\ref{SD23})-(\ref{SD24}) and Eqs.~(\ref{SD27})-(\ref{SD28}) in slit pore geometry}
\label{nums}

We explain here the solution of Eqs.~(\ref{SD23})-(\ref{SD24}) and Eqs.~(\ref{SD27})-(\ref{SD28}) in the specific case of the charges confined to the slit pore. In the slit pore geometry characterized by the plane symmetry with $\e(\br)=\e(z)$ and $\sigma(\br)=\sigma(z)$, the average potentials depend exclusively on the $z$ coordinate, i.e. $\phi_{\rm s,c}(\br)=\phi_{\rm s,c}(z)$. Moreover, the Green's functions satisfy the translational symmetry in the $x-y$ plane, i.e.
\be
\label{sym1}
G_{\rm s,c}(\br,\br')=\int\frac{\mathrm{d}^2\bk}{4\pi^2}\tG_{\rm s,c}(z,z';k)e^{i\bk\cdot(\br-\br')}.
\ee
Using these symmetries, Eqs.~(\ref{SD23})-(\ref{SD24}) take the unidimensional form
\bea
\label{SD29}
&&\frac{k_{\rm B}T}{e^2}\partial_z\e(z)\partial(z)\phi\s(z)-2n_{+\rm b}k_0(z)\sinh\left[\phi\s(z)\right]=-\sigma(z),\nonumber\\
&&\\
\label{SD30}
&&\frac{k_{\rm B}T}{e^2}\left[\partial_z\e(z)\partial_z-\e(z)k^2\right]\tG\s(z,z';k)\nonumber\\
&&-2n_{+\rm b}k_0(z)\cosh\left[\phi\s(z)\right]\tG\s(z,z';k)=-\delta(z-z'),
\eea
with 
\be\label{SD30II}
k_0(z)=e^{-\frac{1}{2}\delta G\s(z)}\theta(z-a)\theta(d-a-z)
\ee
where $a$  stands for the closest approach distance. 

Within the same plane symmetry, the integral Eqs.~(\ref{SD27})-(\ref{SD28}) for the potential components associated with the counterions become equally unidimensional,
\bea
\label{SD31}
\phi\ce(z)&=&-\int_0^d\mathrm{d}z_1\tG\s(z,z_1;k=0)k_0(z_1)\sinh\left[\phi\s(z_1)\right]\nonumber\\
&&\hspace{.5cm}\times\left[q\ce n_{\rm cb}-n_{+\rm b}\delta G\ce(z_1)\right],\\
\label{SD32}
\tG\ce(z,z';k)&=&-\int_0^d\mathrm{d}z_1\tG\s(z,z_1;k)S(z_1)\tG\s(z_1,z';k),
\eea
with
\bea\label{source}
S(z)&=&2n_{+\rm b}k_0(z)\sinh\left[\phi\s(z)\right]\\
&&\times\left\{\phi\ce(z)+q\ce n_{\rm cb}\int_0^d\mathrm{d}z\ce k_{\rm cs}(z\ce)\tG\s(z,z\ce;k=0)\right\}\nonumber\\
&&+\left\{q\ce n_{\rm cb}-n_{+\rm b}\delta G\ce(z)\right\}k_0(z)\cosh\left[\phi\s(z)\right],\nonumber\\
k_{\rm cs}(z)&=&e^{-q\ce\phi\s(z)-\frac{q\ce^2}{2}\delta G\s(z)}\theta(z-a)\theta(d-a-z).
\eea
Finally, taking into account the planar symmetry, the ion densities~(\ref{sp9})-(\ref{sp11}) simplify to
\bea
\label{sp12}
n\ce(z)&=&n_{\rm cb}k_{\rm cs}(z),\\
\label{sp13}
n_{+}(z)&=&n_{+\rm b}k_{+\rm s}(z)\\
&&\hspace{0cm}\times\left\{1-\phi\ce(z)-\frac{1}{2}\delta G\ce(z)+\frac{q\ce n_{\rm cb}}{2n_{+\rm b}}\right.\nonumber\\
&&\hspace{5mm}\left.-q\ce n_{\rm cb}\int_0^d\mathrm{d}z\ce k_{\rm cs}(z\ce)\tG\s(z,z\ce,k=0)\nonumber\right\},\\
\label{sp14}
n_{-}(z)&=&n_{+\rm b}k_{-\rm s}(z)\\
&&\hspace{0cm}\times\left\{1+\phi\ce(z)-\frac{1}{2}\delta G\ce(z)-\frac{q\ce n_{\rm cb}}{2n_{+\rm b}}\right.\nonumber\\
&&\hspace{5mm}\left.+q\ce n_{\rm cb}\int_0^d\mathrm{d}z\ce k_{\rm cs}(z\ce)\tG\s(z,z\ce,k=0)\nonumber\right\}\nonumber\\
&&+q\ce n_{\rm cb}k_{-\rm s}(z),\nonumber
\eea
with $k_{\pm\rm s}(z)=k_0(z)e^{\mp\phi\s(z)}$. We finally note that the numerical results of the main text are obtained from the local ion density functions in Eqs.~(\ref{sp12})-(\ref{sp14}).

\subsubsection{Iterative scheme for the numerical solution of Eqs.~(\ref{SD29}) and~(\ref{SD30})}
\label{sp}

We explain here the iterative scheme for the exact numerical solution of the electrostatic self-consistent Eqs.~(\ref{SD29}) and~(\ref{SD30}) associated with the symmetric monovalent salt. This solution scheme differs from the one introduced in Ref.~\cite{Buyuk2012} where the equations were solved approximatively by truncation. Our scheme is based on the transformation of the differential relation~(\ref{SD30}) into an integral equation. To this aim, we define the Fourier transformed DH-level Green's function 
\be\label{r1}
\left[\partial_z\e(z)\partial_z-\e(z)p^2(z)\right]\tG_0(z,z';k)=-\frac{e^2}{k_{\rm B}T}\delta(z-z'),
\ee
with the screening function $p^2(z)=k^2+\kappa^2_0\theta(z)\theta(d-z)$. From now on, we will omit the dependence of the Fourier transformed functions on the wavevector $k$. First, using Eq.~(\ref{inv}), one can derive from Eq.~(\ref{r1}) the DH-level Green's operator in the form
\be
\label{r3}
\tG_0^{-1}(z,z')=\frac{k_{\rm B}T}{e^2}\left[-\partial_z\e(z)\partial_z+\e(z)p^2(z)\right]\delta(z-z').
\ee
In terms of the operator~(\ref{r3}), Eq.~(\ref{SD30}) can be now expressed as
\bea
\label{r4}
&&\int_{-\infty}^{\infty}\mathrm{d}z_2\tG^{-1}_0(z_1,z_2)\tG\s(z_2,z')+\delta n(z_1)\tG\s(z_1,z')\nonumber\\
&&=\delta(z_1-z'),
\eea
where we defined the excess density function
\be
\label{r5}
\delta n(z)=2n_{+\rm b}\left\{k_0(z)\cosh\left[\phi\s(z)\right]-\theta(z)\theta(d-z)\right\}.
\ee
Finally, multiplying Eq.~(\ref{r4}) with $\tG_0(z,z_1)$, and integrating the result over the variable $z_1$, Eq.~(\ref{SD30}) takes the form of the following integral relation,
\be
\label{r6}
\tG\s(z,z')=\tG_0(z,z')-\int_0^d\mathrm{d}z_1\tG_0(z,z_1)\delta n(z_1)\tG\s(z_1,z').
\ee

In the slit pore geometry, the Fourier-transformed reference Green's function solving Eq.~(\ref{r1}) reads for $0\leq z,z'\leq d$~\cite{Buyuk2012}
\be\label{r7}
\tG_0(z,z')=\tG_{0\rm b}(z,z')+\delta\tG_0(z,z'),
\ee
with the bulk part $\tG_{0\rm b}(z,z')=2\pi\ell_{\rm B}p^{-1}e^{-p|z-z'|}$, and the discontinuous part
\bea\label{r8}
\delta\tG_0(z,z')&=&\frac{2\pi\ell_{\rm B}}{p}\frac{\Delta}{1-\Delta^2e^{-2pd}}\\
&&\times\left\{e^{-p(z+z')}+e^{-p(2d-z-z')}\right.\nonumber\\
&&\hspace{5mm}\left.+2e^{-2pd}\cosh\left[p(z-z')\right]\right\},\nonumber
\eea
where $p=\sqrt{k^2+\kappa_0^2}$ and $\Delta=(\e_{\rm w}p-\e_{\rm m}k)/(\e_{\rm w}p+\e_{\rm m}k)$. Finally, the ionic self-energy associated with the Green's functions~(\ref{r6}) and~(\ref{r7}) should be obtained from the numerical evaluation of the Fourier integral 
\be
\label{r9}
\delta G_\alpha(z)=\int_0^\infty\frac{\mathrm{d}kk}{2\pi}\delta\tG_\alpha(z,z)
\ee
for $\alpha=\{0,{\rm s}\}$.

We explain now the cyclic solution of Eqs.~(\ref{SD29}) and~(\ref{r6}) for the average potential and Green's function associated with the monovalent salt. First, Eq.~(\ref{SD29}) should be numerically solved on a discrete lattice by approximating the self-energy $\delta G_{\rm s}(z)$ inside the function $k_0(z)$ in Eq.~(\ref{SD30II}) with the DH self-energy $\delta G_0(z)$ in Eq.~(\ref{r9}). This solution should satisfy the following boundary conditions associated with the surface charge distribution in Eq.~(\ref{dench2}),
\be
\left.\phi'\s(z)\right|_{z=0}=4\pi\ell_{\rm B}\sigma\s\;;\hspace{3mm}\left.\phi'\s(z)\right|_{z=d}=-4\pi\ell_{\rm B}\sigma\s.
\ee
Then, the output potential $\phi\s(z)$ should be used in Eq.~(\ref{r5}) for the iterative solution of the integral Eq.~(\ref{r6}). The latter will be called the \textit{inner cycle}. At the first step of the inner cycle, the r.h.s. of Eq.~(\ref{r6}) should be evaluated by replacing the unkown Green's function $\tG\s(z,z')$ by the reference Green's function $\tG_0(z,z')$. At the second step, the resulting output Green's function $\tG\s(z,z')$ should be inserted into the integral on the r.h.s., and the inner cycle should be iterated until numerical convergence is achieved. Finally, at the end of the inner cycle, Eq.~(\ref{SD29}) should be solved again with the updated value of the ionic self energy $\delta G\s(z)$ in Eq.~(\ref{SD30II}), and the outer cycle composed of the iterative solution of Eqs.~(\ref{SD29}) and~(\ref{r6}) should be continued along the same lines until the solution stabilizes.

\subsubsection{Iterative solution of the integral Eqs.~(\ref{SD31}) and~(\ref{SD32})}

Finally, we explain here the iterative solution of the integral Eqs.~(\ref{SD31})-(\ref{SD32}). These equations include the counterion contribution to the self energy defined by Eq.~(\ref{sp6II}). In the plane geometry, this potential reads
\be\label{r10}
\delta G\ce(z)=\int_0^\infty\frac{\mathrm{d}kk}{2\pi}\left[\tG\ce(z,z)-\tG_{\rm cb}(0)\right].
\ee
The bulk Green's function inside the integral of Eq.~(\ref{r10}) can be easily computed by solving Eq.~(\ref{SD26}) in the bulk limit where $\phi\s(z)=0$, $\delta G\ce(z)=0$, $k_0(z)=1$, and $S(r)=q\ce n_{\rm cb}$. This trivial calculation gives $\tG_{\rm cb}(0)=-q\ce n_{\rm cb}(2\pi\ell_{\rm B})^2/p^3$. 

The solution of Eqs.~(\ref{SD31}) and~(\ref{SD32}) requires the knowledge of the potentials $\phi\s(z)$ and $\tG\s(z,z')$ associated with the salt solution. The numerical computation of these potentials was explained in Appendix~\ref{sp}. After these potentials have been computed, the first step of the solution scheme consists in evaluating the r.h.s. of Eq.~(\ref{SD31}) by neglecting the self-energy component $\delta G\ce(z)$ inside the integral. At the second step, the output potential $\phi\ce(z)$ is used to evaluate the r.h.s. of Eq.~(\ref{SD32}) by neglecting the self-energy $\delta G\ce(z)$ inside the auxiliary function $S(z)$. Next, the resulting Green's function $\tG\ce(z,z')$ should be inserted into Eq.~(\ref{r10}) for the evaluation of the the counterion contribution to the self energy $\delta G\ce(z)$. The latter should be finally used for the updated solution of Eq.~(\ref{SD31}), and this cycle should continue along the same lines until numerical convergence is achieved.


\begin{thebibliography}{99}
\bibitem {foot1} The truncation of the structure function $F(\br,\br\ce)$ at the order $O\left(\Lambda^0\ce\right)$ is justified by the fact that this function is already multiplied in Eq.~(\ref{SD6}) by the multivalent counterion density $n\ce(\br\ce)$.
\end{thebibliography}
\end{document}